\documentclass[preprint,aps,showpacs,preprintnumbers,amsmath,amssymb,nofootinbib]
{revtex4}

\usepackage{graphicx}
\usepackage{epsfig}
\begin{document}

\newcommand{\mn}{\Delta m_{21}^2}
\newcommand{\mt}{\Delta m_{31}^2}
\newcommand{\si}{s_{12}}
\newcommand{\sn}{s_{23}}
\newcommand{\st}{s_{13}}
\newcommand{\ci}{c_{12}}
\newcommand{\cn}{c_{23}}
\newcommand{\ct}{c_{13}}
\newcommand{\nova}{NO$\nu$A~}
\newcommand{\dcp}{\delta_{\rm CP}}

\title{Revisiting the sensitivity studies for leptonic CP violation and mass hierarchy with T2K,
NO$\nu$A and LBNE experiments}

\author{K. N. Deepthi, Soumya C.,  R. Mohanta}
\affiliation{
School of Physics,  University of Hyderabad, Hyderabad - 500046,  India }

\begin{abstract}
 Precision measurement of the neutrino mixing parameters and the determination of mass hierarchy are
the primary goals of the present and upcoming neutrino experiments.  In this work, we study the sensitivity of T2K,
NO$\nu$A and LBNE experiments to discover leptonic CP violation and the determination of neutrino mass hierarchy.
We  obtain the correlation between the CP violating phase $\delta_{CP}$ and the mixing angles $\theta_{13}$, $\theta_{23}$
and the sensitivity  to determine the octant of atmospheric mixing angle $\theta_{23}$. The entire analysis is done
for a total 10 years  (5$\nu$+ 5$\bar \nu$)   of running of T2K,  NO$\nu$A and LBNE experiments. Furthermore, we also 
consider the impact of cross section uncertainties on the CP violation sensitivity of LBNE experiment.  \\

\end{abstract}


\maketitle

\section{Introduction}
\label{introduction}

The experimental endeavour in the past decades has firmly established the phenomenon of
neutrino oscillations, i.e., the composition of neutrino flavors change as they propagate.
In the three neutrino framework, the three flavour eigenstates ($\nu_e, \nu_\mu, \nu_\tau$)
 mix via the unitary lepton mixing matrix $U_{PMNS}$ \cite{pmns,pmns-a, valle-1},   analogue of the CKM matrix
$V_{CKM}$ that governs the  mixing in the quark sector. This PMNS matrix
can be parameterized in terms of three mixing angles ($\theta_{12}, \theta_{13}$  and $\theta_{23}$),
which have all been measured experimentally and a CP-violating
phase $\delta_{CP}$ which is unknown. The probability for flavor oscillation also depends on the
differences in the squared masses of the neutrinos, i.e., $\Delta m_{21}^2$ and $\Delta m_{31}^2$,
where $\Delta m_{ij}^2 = m_i^2 - m_j^2$.

The neutrino oscillation data accumulated over many years allows us to determine the solar
and atmospheric neutrino oscillation parameters with very high precision.
The mixing angles $\theta_{12}$ and $\theta_{23}$ as well as the mass square differences have
been well constrained by various neutrino experiments.
Recently, the reactor mixing angle $\theta_{13}$  has been  measured precisely \cite{daya-bay,daya-bay1,reno,t2k}
with a moderately large value.
This provides a significant achievement in establishing the picture of three-flavor
neutrino oscillations. The global analysis of the recent results of various neutrino oscillation experiments
has been performed by several groups \cite{gfit1,gfit2,gfit3,gfit4}. We have considered best-fit values and the $3\sigma$ ranges of
the oscillation parameters from Ref. \cite{gfit4} through out in our simulations.

There are however, still many open questions to be answered. These include :
i.) The value of the CP violating phase $\delta_{CP}$ is not yet constrained by any experiment.
ii.) We still do not know the exact nature of neutrino mass hierarchy, i.e., whether the
neutrino mass ordering is normal or inverted in nature.
iii.) The possibility
of observing CP violation in the neutrino sector due to the
presence of the Dirac type CP violating phase in the neutrino mixing matrix.
iv.) Another interesting and crucial development in recent times  is the
indication of non-maximal atmospheric mixing angle by the MINOS \cite{minos} and T2K \cite{t2k-1} experiments.
The global analyses of all the available neutrino oscillation data
 \cite{gfit1, gfit2, gfit3, gfit4} also prefer the deviation of $\theta_{23}$ value from maximal mixing
 i.e., $\sin^2 \theta_{23}  \neq  0.5 $. Thus, for non-maximal value of $\theta_{23}$, one can have two
possible solutions, one with  $ \theta_{23}< 45^\circ$ for which $( \sin^2 \theta_{23}-0.5)$ is negative
and the other with $\theta_{23}>45^\circ$ for which $(\sin^2 \theta_{23}-0.5)$ is positive. The former case is known as 
lower octant (LO) whereas the later one is
known as higher octant (HO) solution. This corresponds
to the problem of  octant degeneracy of $\theta_{23}$.
 In this paper we would like to  study the sensitivity of the current
and future long baseline experiments i.e., T2K, NO$\nu$A and LBNE in addressing  some of these issues.
Although some of these aspects have been studied in detail recently  by various authors
\cite{Agarwalla:2013ju,Chatterjee:2013qus,sruba1, sruba2, sruba3,gandhi,Hiraide:2006vh,bora,minakata-1,minakata-2},
in this paper we have attempted to do
a complete analysis of all these issues in the context of the current generation and upcoming long baseline super-beam
experiments. Another important difference is that in most of the previous analyses the LBNE flux files used are
either atmospheric or NO$\nu$A (which is an off-axis experiment) flux files whereas we have considered the on-axis
NuMI beam flux files for LBNE from \cite{lbneflux}. In Ref. \cite{gandhi}, the authors have studied the sensitivities to mass
 hierarchy, octant of $\theta_{23}$ and CP violation for LBNE. They have also included the data from T2K (5+0), NO$\nu$A (3+3)
 and also from atmospheric neutrinos. The difference between their work and ours are: (i) we have not taken into account the effect
 atmospheric neutrinos (ii) we have considered 10 years of data for No$\nu$A and T2K in the combinations (5+5) assuming that by
 the time LBNE will start data taking both NO$\nu$A and T2K would have completed 10 years of run (iii) we have also studied
 various correlations between $\delta_{CP}$ and $\theta_{23}/\theta_{13}$, which will help us to constrain the value of $\delta_{CP}$.
Furthermore, as discussed in Ref. \cite{x-section}, the uncertainties in cross-sections play a crucial role in the determination
of CP violation sensitivities of various long baseline super-beam experiments. Without considering any specific theoretical model, 
the errors on cross-sections are expected to be in the range of (20-50)\%. In this paper, we have studied the impact of these 
cross-section uncertainties on the CP violation sensitivity of LBNE experiment.

The paper is organized as follows. In Section 2, we discuss the $\delta_{CP}$ dependence of neutrino
oscillation probabilities   and also show how it is correlated with
the octant of $\theta_{23}$ and neutrino mass ordering. The experimental details of
the long-baseline experiments (NO$\nu$A, T2K and LBNE) are briefly discussed in Section 3.
The CP violation sensitivity and the determination of mass hierarchy are outlined in sections 4 and 5.
Section 6 contains the  results on octant sensitivity determination of these experiments. The correlations between
the CP violating phase $\delta_{CP}$ and the mixing angles $\theta_{12}$ and $\theta_{23}$ are presented in
Section 7. Section 8 contains the summary and conclusion.

\section{Effect of Mass Hierarchy and $\theta_{23}$ octant on $\delta_{CP}$ Sensitivity}

The three-flavor neutrino oscillation effects can be systematically demonstrated by considering oscillation channels $\nu_\mu
\to \nu_e $ and $\bar{ \nu}_\mu \to \bar{\nu}_e$. The detailed study of these  channels at the  long-baseline experiments is
capable of addressing almost all the  four major issues discussed in the previous section. In particular,
 appearance channel, i.e., $ \nu_\mu
\to \nu_e $ is very sensitive to explore the CP violation effect in neutrino oscillation experiments which can be understood
as follows. In matter of constant density, the appearance  probability $P_{\mu e}$, which  depends on $\delta_{CP}$  in its sub-leading
term can be expressed  as \cite{prob1,prob2,prob3}
\begin{eqnarray}
&&P_{\mu e}  \simeq  \sin^2 \theta_{23} \sin^2 2 \theta_{13} \frac{\sin^2[(1-\hat A) \Delta]}{(1-\hat A)^2}\nonumber\\
&&~~+ \alpha \sin 2 \theta_{13} \sin 2 \theta_{12} \sin 2 \theta_{23} \cos \theta_{13}
\cos(\Delta+\delta_{CP}) \frac{\sin (\hat A \Delta)}{\hat A}\frac{\sin[(1-\hat A) \Delta]}{1-\hat A}\;,\label{eq1}
\end{eqnarray}
where $\alpha = \Delta m_{21}^2/\Delta m_{31}^2$,  $\Delta\equiv \Delta m_{31}^2L/4E$,
$\hat A = 2 \sqrt 2 G_F N_e E/\Delta m_{31}^2$.
All six parameters governing neutrino oscillations $(\theta_{12},~ \theta_{23},~ \theta_{13}, ~\Delta m^2_{21},~\Delta m_{31}^2~{\rm and} ~
\delta_{CP})$ appear in this equation. It should be noted that the parameters
$\alpha$, $\Delta$ and $\hat A$ are sensitive to the neutrino mass ordering i.e.,
to the sign of $\Delta m_{31}^2$. Furthermore, the sign of $\hat A$ changes with the sign of
$\Delta m_{31}^2$, which implies that the matter effect can be used to determine the
mass hierarchy. Also $\hat A$ changes sign while going from neutrino to antineutrino mode, which
indicates that it can mimic CP violation and hence complicates the extraction of $\delta_{CP}$
by comparing the data from neutrino and antineutrino modes. Thus, for large $\theta_{13}$ from the
dominant first term of Eq. (1), one can determine $\sin^2 \theta_{23}$  or in other words the octant
of $\theta_{23}$. Secondly,  as this term contains large matter effect, the nature of mass ordering can also
be extracted from it. The second  sub-dominant term
is sensitive for the determination of CP violation  as  it contains both $\sin \delta_{CP} $ and $\cos \delta_{CP} $ terms.
As discussed in detail in Ref. \cite{sruba1}, the following points can be inferred from Eq. (1).

$\bullet$  The CP violation phase  $\delta_{CP}$  appears in combination with the atmospheric mass-squared difference as
$\cos(\Delta +\delta_{CP} )$ and hence,  it suffers from the hierarchy-$\delta_{CP}$ degeneracy \cite{minakata-1}. This in turn  limits
the CP violation sensitivity which can be clearly understood from Fig.-1, where we have plotted the $P_{\mu e}$ energy spectrum
for LBNE experiment which has baseline of 1300 km. 
In our analysis, we have used the atmospheric oscillation parameters $\Delta m_{atm}^2$ and $\theta_{\mu \mu}$ extracted from 
the muon neutrino survival probability. For the muon neutrino disappearance experiment such as MINOS \cite{minos-5}, T2K
 \cite{t2k-5} and NO$\nu$A  the survival
probability is given by
\begin{equation}
P_{\mu \mu}=1 - \sin^2 \theta_{\mu \mu} \sin^2 \left (\frac{\Delta m_{atm}^2 L}{4E} \right )+ {\cal O} (\Delta m_{21}^2)\;.
\end{equation} 
The relation between the atmospheric parameters
 measured and the standard three flavour  oscillation parameters are given as \cite{rel,rel-1,rel-2}\\
\begin{equation}
\sin\theta_{23} =\frac{\sin\theta_{\mu\mu}}{\cos\theta_{13}}\\
\end{equation}
\begin{equation}
\Delta m_{31}^2 = \Delta m_{atm}^2 +\Delta m_{21}^2 (\cos^2 \theta_{12}- \cos \delta_{CP} \sin \theta_{13} \sin 2\theta_{12} \tan \theta_{23})\\
\end{equation}
 where $\Delta m_{atm}^2$ is taken to be positive (negative)  for Normal Hierarchy (Inverted Hierarchy). These relations become significant
in the light of the recently measured moderately large value of $\theta_{13}$. Therefore, in order to avoid the erroneous results for the
sensitivity studies, we use these corrected atmospheric parameters in our analysis.

 We consider the true curves of $\delta_{CP}=\pm 90^\circ$ and true
hierarchy to be normal for both the panels. The test values for $\delta_{CP}=0$ and $180^\circ$ and test NH is shown in left
panel and the same for test hierarchy as inverted is shown on the right panel. Thus, the left panel represents the separation
between the CP conserving test ($\delta_{CP}^{test}= 0, \pi$) and maximally CP violation true  ($\delta_{CP}^{true}= -\pi/2 ~{\rm or}~\pi/2$)
when the hierarchy is known while the right panel represents the same when the hierarchy is unknown. Hence, one can see that
the separation between the true cases i.e., (NH, $\delta_{CP}=\pm \pi/2$) from the corresponding test CP conserving
cases (NH/IH, $\delta_{CP}$=0 or $\pi$) is hierarchy dependent, which will effectively introduce hindrance in the CP
sensitivity measurements.

$\bullet$  The probability  for neutrinos $P_{\mu e}$ is higher
for NH than for IH due to matter effects as seen from the first term in Eq. (1).

$\bullet $ The second term of Eq. (1), which  is sensitive to  $\delta_{CP}$ gives rise to
 intrinsic octant degeneracy \cite{lisi}, as it comes  with $\sin 2 \theta_{23}$ term, i.e., $ P_{\mu e} (\theta_{23} ) = P_{\mu e} (\pi/2  - \theta_{23} )$.
Furthermore, it has 
been shown in Refs. \cite{barger-11, minakata-11} that the probability function for different values of $\theta_{13}$ and $\delta_{CP}$ may satisfy the
relation $P_{\mu e}(\theta_{23}^{true}, \theta_{13}, \delta_{CP})=P(\theta_{23}^{wrong}, \theta_{13}', \delta_{CP}')$
inducing the overall eight-fold degeneracy.
This implies that there could be some probability  that the test values of $\theta_{23}$ occurring anywhere
in the `wrong' octant may give the same probability.
The effect of octant degeneracy in distinguishing between the CP conserving ($\delta_{CP}=0$ or $\pi$) and maximally
CP violating cases ($\delta_{CP}= \pm \pi/2$) for $P_{\mu e}$ are depicted in Fig.-2, where the upper panel is for neutrinos and lower one for
anti-neutrinos. The shaded bands correspond to the true  value of $\theta_{23}$ in the lower octant (LO). The
figure on the top left panel  shows that
for the true LO and true $\delta_{CP}=-\pi/2$, the true  case can't be distinguished whereas for the plot in the right
panel for $\delta_{CP}=\pi/2$, there exists  a clear distinction. For anti-neutrinos the behavior of $\delta_{CP}=-\pi/2$ and
$\delta_{CP}=\pi/2$ is opposite.
This fact implies that the combination $\nu$'s and $\bar{\nu}$'s would be well suited  for the removal of octant-$\delta_{CP}$ degeneracy.
Furthermore, recently it has been pointed out in Ref. \cite{coloma} that, the CP violating phase 
$\delta_{CP}$  and the mixing angle $\theta_{23}$  can be measured precisely 
in an environment where there are strong correlations between them. 
This can be achieved by paying special attention to  
the appearance and the disappearance channels in long-baseline neutrino oscillation experiments. 

\begin{figure}[htb]
\includegraphics[width=8cm,height=7cm, clip]{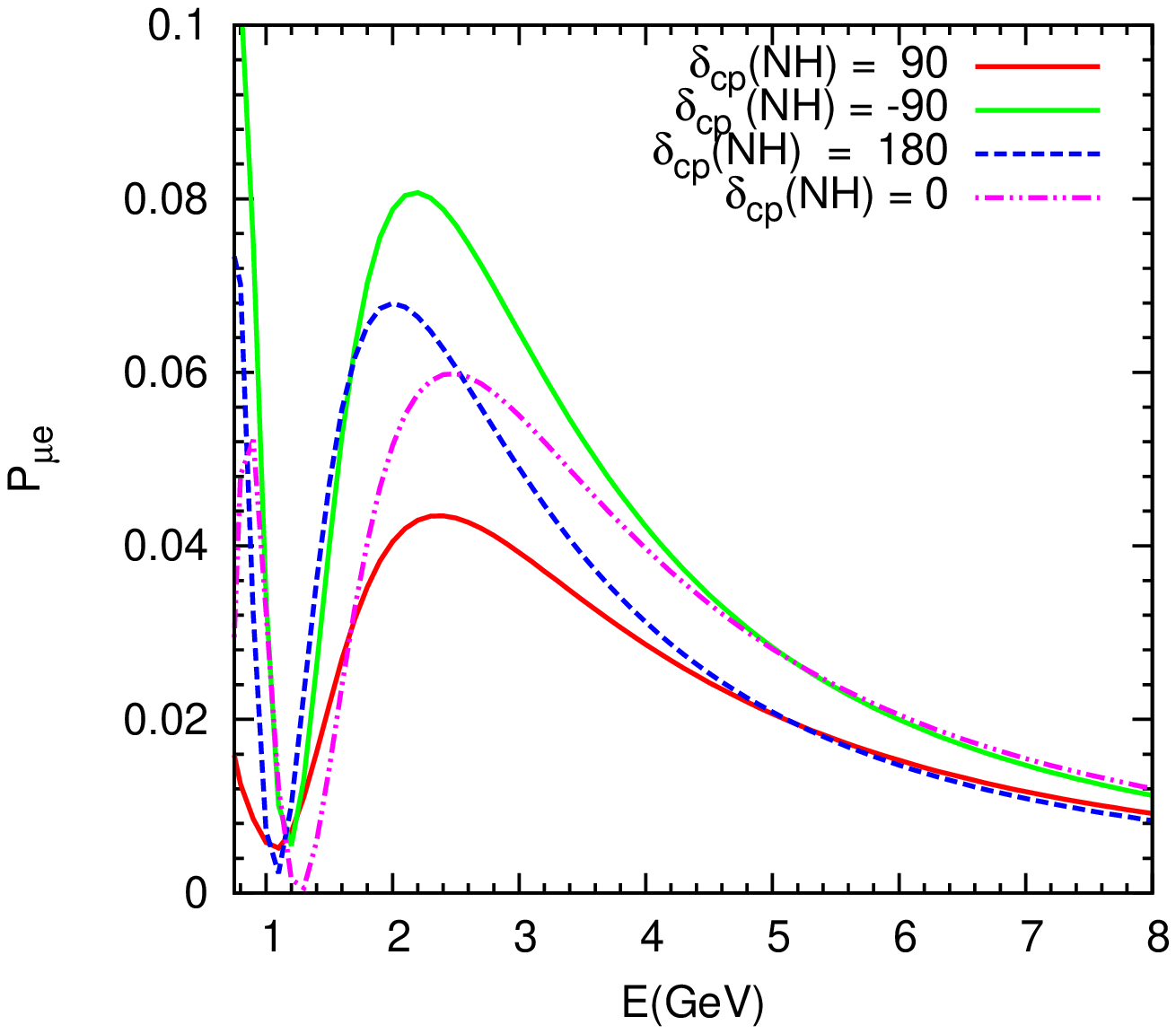}
\hspace{0.2 cm}
\includegraphics[width=8cm,height=7cm, clip]{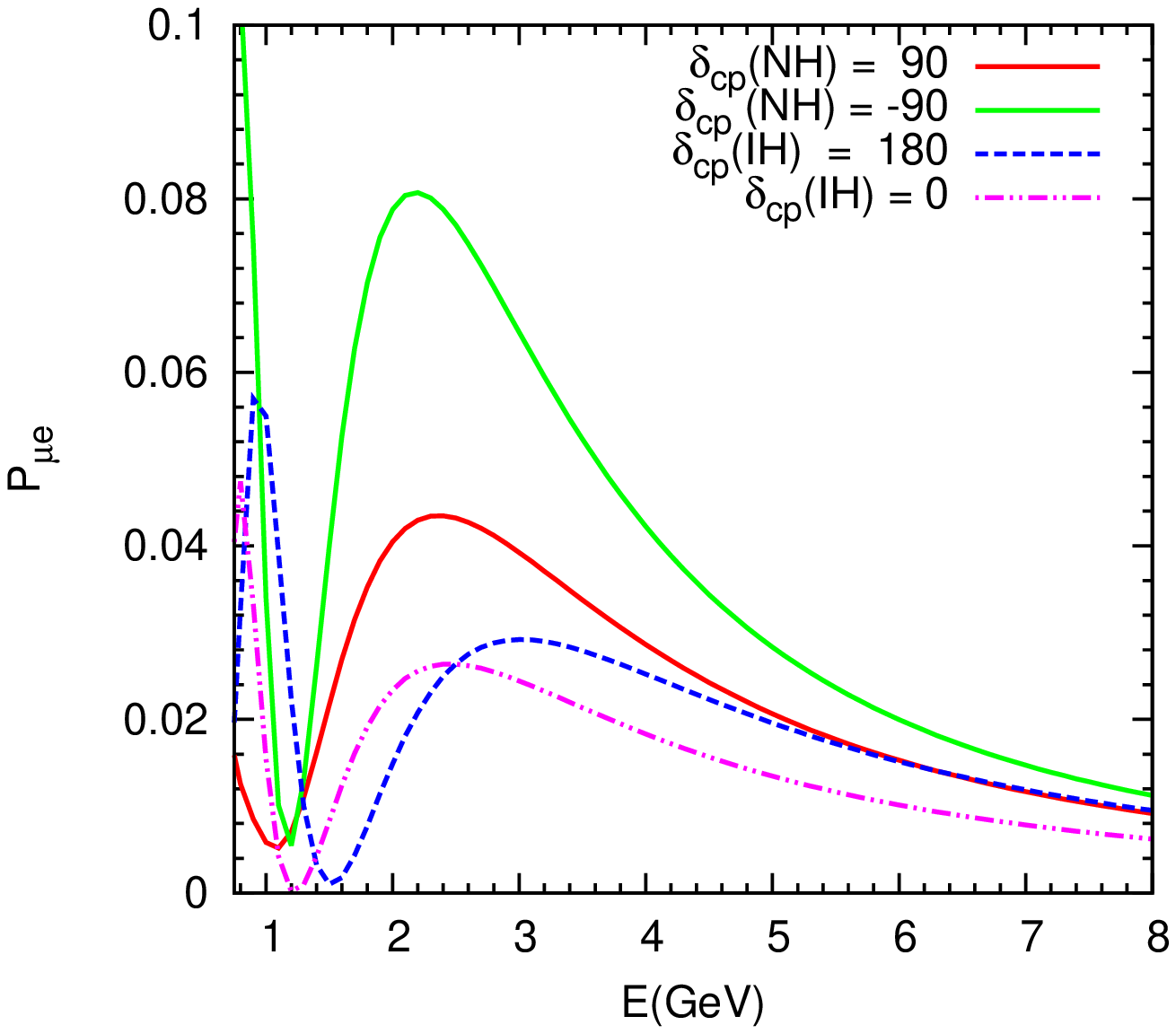}
\caption{The  energy spectrum $P_{\mu e}$ for true $\delta_{CP}= \pm 90^\circ$ and  true hierarchy as NH (both panels),
test $\delta_{CP}=0$ and $180^\circ$ for NH (left panel) and IH (right panel). Here we have used $\sin^2\theta_{23}=0.41$, $\sin^2 2 \theta_{13}=0.1$
and baseline $L=1300$ km for LBNE experiment.}
\end{figure}

\begin{figure}[!h]
\begin{centering}
\begin{tabular}{cc}
\includegraphics[width=0.49\columnwidth]{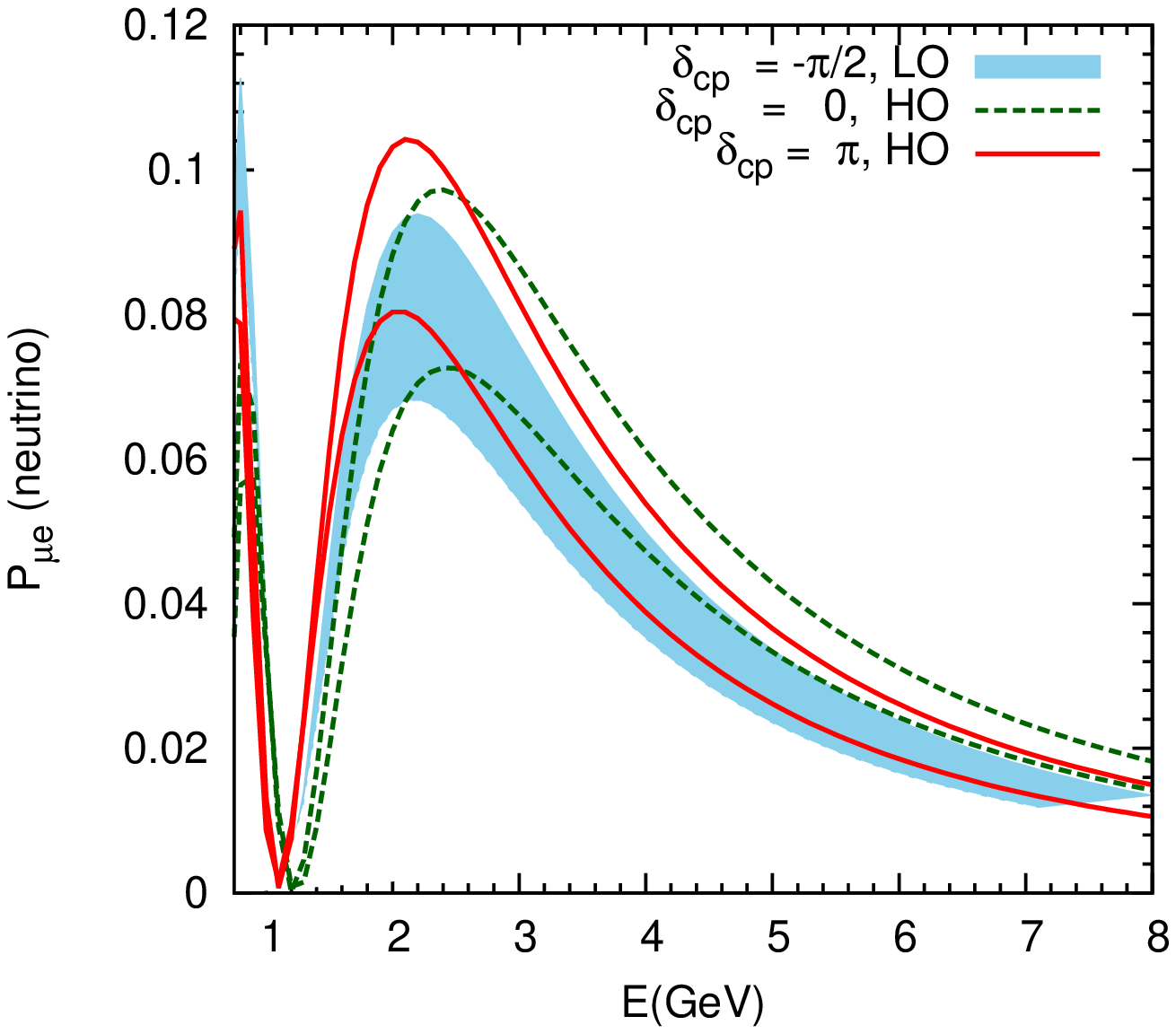}
\includegraphics[width=0.49\columnwidth]{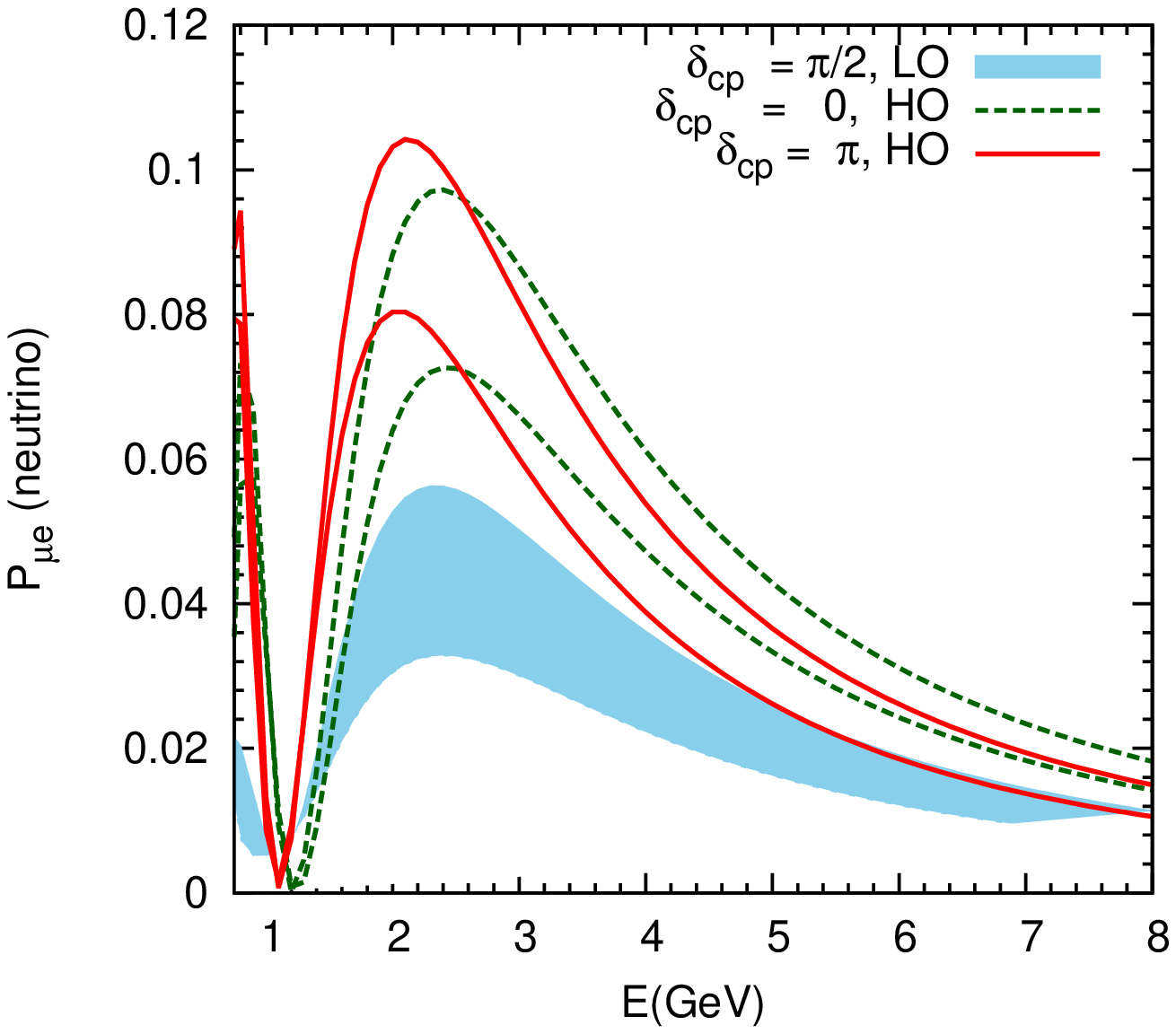}\\
\includegraphics[width=0.5\columnwidth]{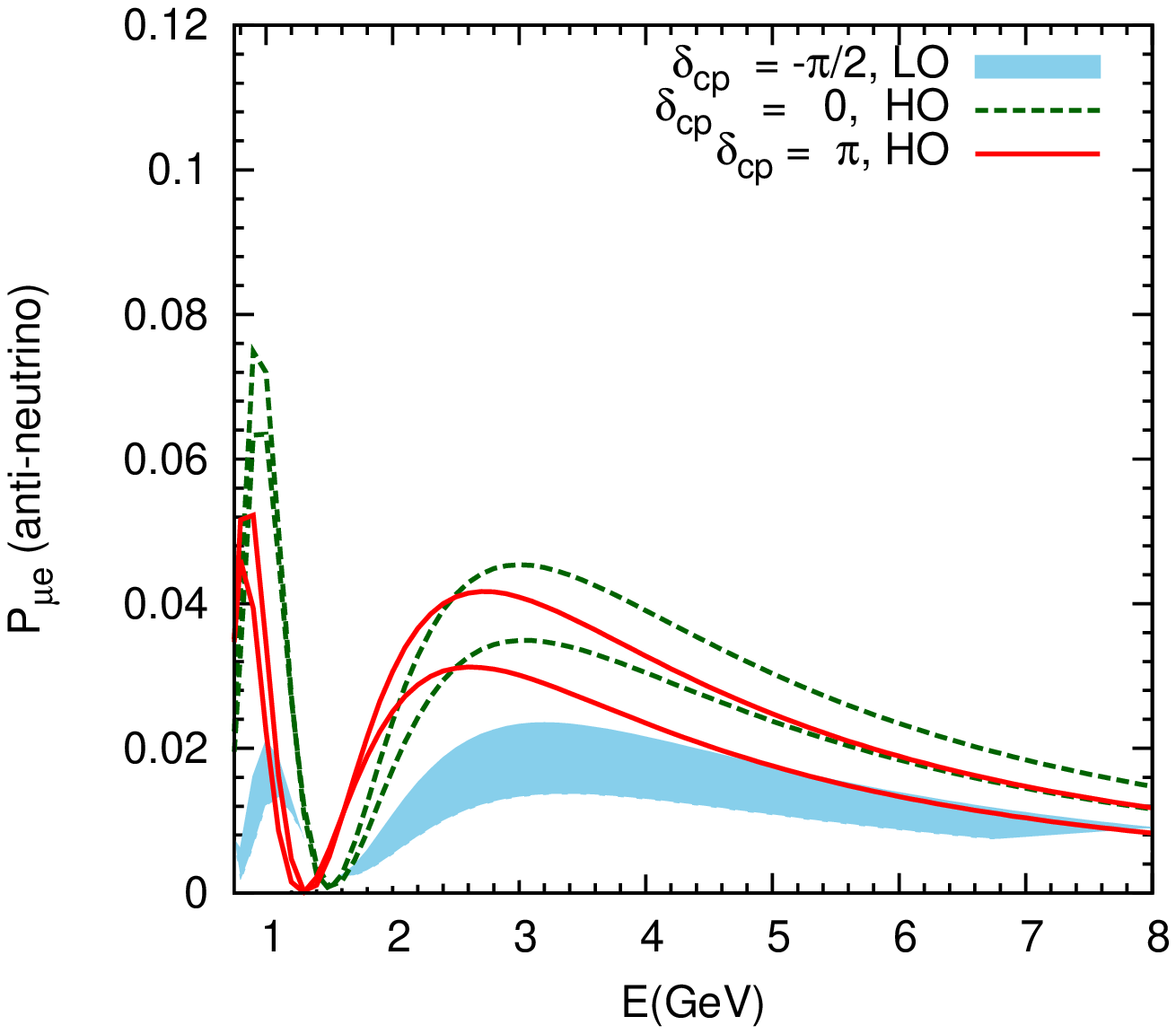}
\includegraphics[width=0.5\columnwidth]{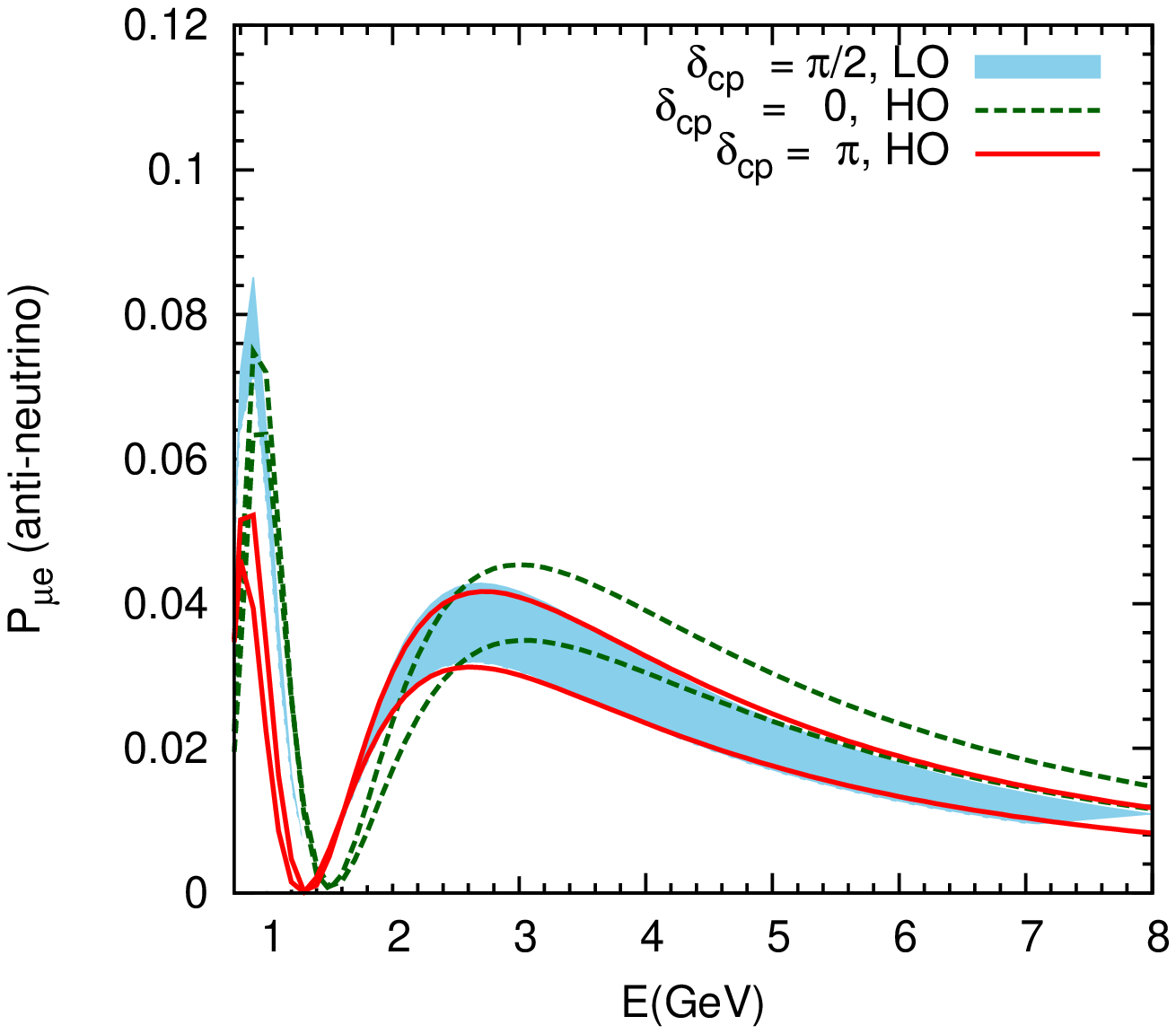}
\end{tabular}
\par\end{centering}
\caption{$P_{\mu e}$ energy spectrum for LBNE, which demonstrates the octant-$\delta_{CP}$ degeneracy. The upper panels (lower panels)
are for $\nu$ ($\bar \nu$).}

\end{figure}
Gathering information about  these observational facts, we would now proceed to study the sensitivities of various observables in the current
and upcoming long-baseline experiments e.g., NO$\nu$A, T2K and LBNE.

\section{Experimental Specifications for the simulation studies}

To determine the sensitivity  of various observables in the currently running and upcoming long-baseline experiments,  the  simulation is done using the GLoBES package
\cite{Huber:2004gg,Huber:2009xx}. First we briefly describe the procedure that we have adopted for obtaining the numerical results. We calculate $\Delta \chi^2$
using the default definition in GLoBES. We  then minimize the $\Delta \chi^2$ to compute the sensitivities on various parameters.
The following are the  experimental specifications for T2K, NO$\nu$A and LBNE setups that have been used in
our analysis.

\textbf{ T2K: } In the T2K experiment, a $\nu_\mu$ beam from J-PARK is directed towards Super-Kamiokande
detector which is 22.5 kt (Water Cerenkov detector), 295 km away. It uses a 0.77 MW beam planned to run effectively for 5($\nu$) + 0($\bar \nu$) or 3($\nu$) + 2($\bar \nu$)
years. The initial plan of
T2K experiment was to run for five years  with $10^{21}$ proton on target per year.
In this paper we consider the option of T2K running for 5($\nu$) + 5($\bar \nu$)
years  and incorporate those results with NO$\nu$A and LBNE 10 years of run. The details of T2K experiment can be found
from \cite{t2k-3}. We have considered the input files for T2K from GLoBES package alongwith the inputs from \cite{t2k-2,t2k-3,t2k-4}.

\textbf{ NO$\nu$A: } NO$\nu$A is a 14 kt totally active scintillator detector (TASD) located at Ash River, a distance of
810 km from Fermilab \cite{nova,nova-1}.  The beam power is assumed to be 0.7 MW NuMI beam with  $6.0   \times 10^{20}$ pot/year.
This experiment is scheduled  to have three years in neutrino mode first and after that three years run in anti-neutrino mode.
However, in our analysis we consider the running
for  5($\nu$) + 5($\bar \nu$) years by 2024.  \\
The following are the signal and background efficiencies considered in our simulation: \\
 Signal efficiency : 45\% for $\nu_{e}$ and $\bar{\nu}_{e}$ signal; 100\% $\nu_{\mu}$ CC and $\bar{\nu}_{\mu}$ CC.\\
Background efficiency : \\
a) Mis-ID muons acceptance : 0.83\% $\nu_{\mu}$ CC, 0.22\% $\bar{\nu}_{\mu}$ CC; \\
b) NC background acceptance : 2\% (3\%) $\nu_{\mu}$($\bar{\nu}_{\mu}$) NC; \\
c) Intrinsic beam contamination : 26\% (18\% ) $\nu_e$ ($\bar{\nu}_{e}$). \\
We consider $5 \%$ uncertainty on signal normalization and $10 \%$ on background normalization.
The migration matrices for NC background smearing are taken from \cite{sanjib}.

\textbf{ LBNE:}
For LBNE, we consider 35 kt LAr detector at 1300 km baseline length \cite{lbne}.
The neutrino beam (0.5 - 8 GeV) is obtained from a proton beam of 700 KW beam power and 120 GeV beam energy resulting in
$6 \times 10^{20} $ protons on target (POT) per year.
We consider 5 years of data taken by detector  in $\nu$ beam mode and
5 years in $\bar{\nu} $ beam mode. The GLoBES files and the detector parameter assumptions are taken from \cite{lbne-2}.
We consider $5 \%$ uncertainty on signal normalization and $10 \%$ on background normalization.
Furthermore, we have not considered the effect of near detector (ND) in our analysis. As discussed in Ref. \cite{gandhi},
the presence of ND will reduce the systematic uncertainties of $\nu_e$ signal (background)
from $5\%$ (10$\%$) to $1\%$ (1$\%$) and this in turn will enhance the various sensitivities a bit more.

Our primary objective is to perform the sensitivity studies with LBNE setup. However, by the time LBNE
will start data taking, which is expected to be around 2022, both T2K and NO$\nu$A would have completed their 
scheduled run period. However, if they continue data taking beyond their planned run periods, it is interesting to 
study what would be the additional 
information that we can achieve by combining the data from different experiments.
Therefore, we would incorporate the T2K and NO$\nu$A data to the LBNE data set to perform
the simulation. For all the three experiments we consider two cases of runs in neutrino and anti-neutrino modes:
i) 5 yrs in neutrino mode and 5 yrs in anti-neutrino mode.

\section{CP Violation sensitivity with T2K, NO$\nu$A and LBNE}

The determination of the CP violating phase $\delta_{CP}$ is one of the most challenging problems in neutrino physics today.
Since $\delta_{CP}$ is associated  with the mixing angle $\theta_{13}$ in the PMNS matrix, the recent measurement of
a non-zero and moderately large value of this angle by reactor and accelerator experiments is
expected to be conducive for the measurement of $\delta_{CP}$.
Since $\theta_{13}$ is found to be moderately large  it is
possible for NO$\nu$A and T2K to provide some hint on $\delta_{CP}$. In this section, we
discuss the detection of CP violation, i.e., the ability of an experiment to exclude the cases
$\delta_{CP}$ = 0 or $180^\circ$

The sensitivity of the experiment to observe CP violation is evaluated at a given value of $\delta_{CP}$
by minimizing the $\chi^{2}$ at the fixed test values of 0 and $\pi$. Thus, we determine two quantities:
\begin{equation}
\Delta \chi_{0}^{2}=\chi^{2}(\delta_{CP}=0)-\chi^{2}_{true} \nonumber
\end{equation}
\begin{equation}
\Delta \chi_{\pi}^{2}=\chi^{2}(\delta_{CP}=\pi)-\chi^{2}_{true}
\end{equation}
and then take
\begin{equation}
\Delta \chi^{2} = {\rm min}(\Delta \chi_{0}^{2},~\Delta \chi_{\pi}^{2})
\end{equation}
The significance of CP violation is obtained by using $\sigma = \sqrt{\Delta \chi^{2}}$. 
The true values for different oscillation parameters considered in this analysis are:
\begin{eqnarray}
&& \sin^2 \theta_{12}=0.32,~~~~~~\sin^2 2 \theta_{13}=0.1,~~~~~\sin^2 \theta_{23}=0.5,~~~~ \delta_{CP}= [-\pi:\pi],\nonumber\\
&&\Delta m_{21}^2 = 7.6 \times 10^{-5}~{\rm eV^2},~~~~~~\Delta m^2_{atm}=\pm 2.4 \times 10^{-3}~{\rm eV^2~for ~NH/IH}. \label{true-values}
 \end{eqnarray}
Furthermore, we have done marginalization over both the hierarchies, 
 $\sin^2 \theta_{23}$, $\sin^2 2 \theta_{13}$, within the following ranges:  $|\Delta m_{atm}^2| \in [2.05:2.75]\times 10^{-3}~{\rm eV}^2$, 
$\sin^2 \theta_{23} \in [0.32 , 0.68]$,
$\sin^2 2 \theta_{13} \in [0.07:0.13]$. We also added prior for $\sin^22\theta_{13}$ with $\sigma(\sin^22\theta_{13}) =0.01$.
We present our results as a function of $\delta_{CP}$ in Fig.-3.
The 3$\sigma$ (5$\sigma$) line corresponds to
$\Delta \chi^{2}$ = 9 (25) (this correspondence is for 1 degree of freedom), which indicates $99.73\%$ ($99.99\%$) probability of determining the CP violation.
One can notice from the figure that  NO$\nu$A and T2K suffer from the hierarchy-$\delta_{CP}$
degeneracy, because of which their CP detection potential is compromised for unfavorable
values of $\delta_{CP}$. This degeneracy can be lifted by including information from LBNE, which
excludes the wrong hierarchy solution.
\begin{figure}[!ht]
\begin{center}
\includegraphics[width=0.49\columnwidth]{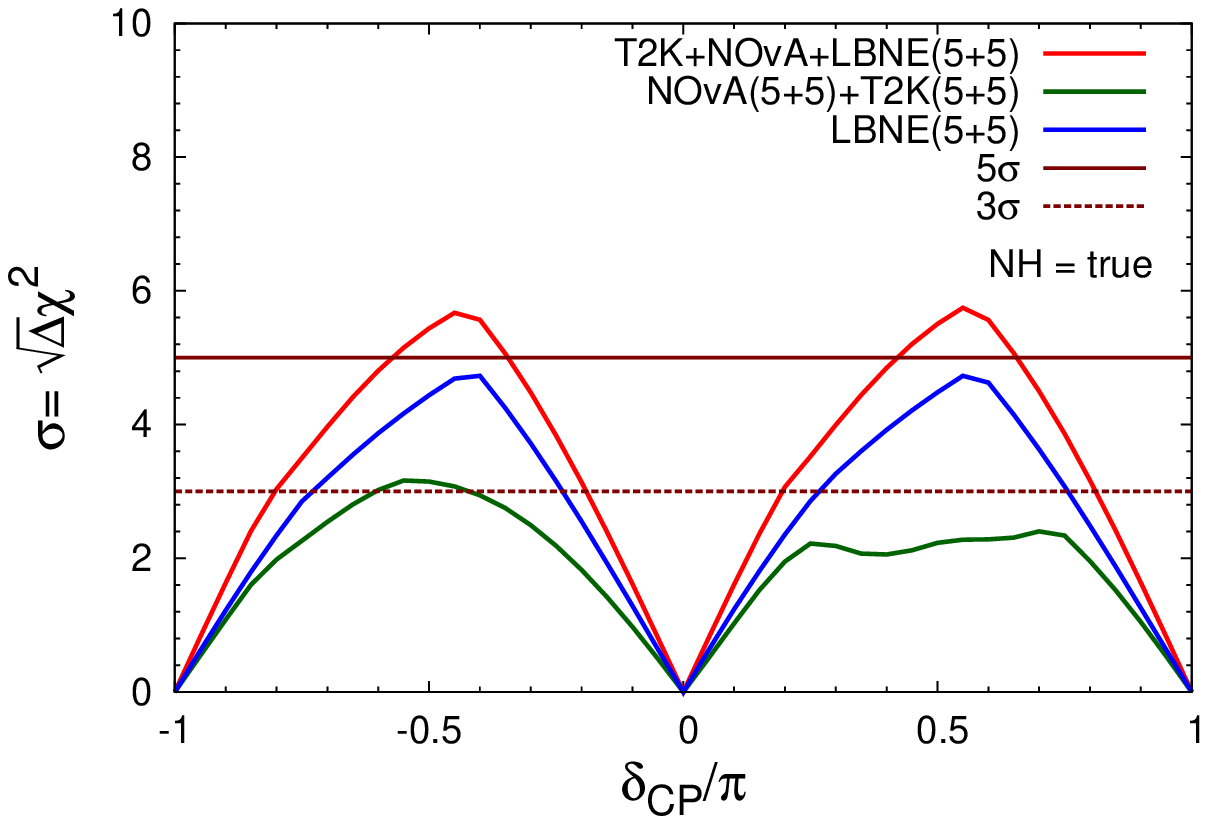}
\includegraphics[width=0.5\columnwidth]{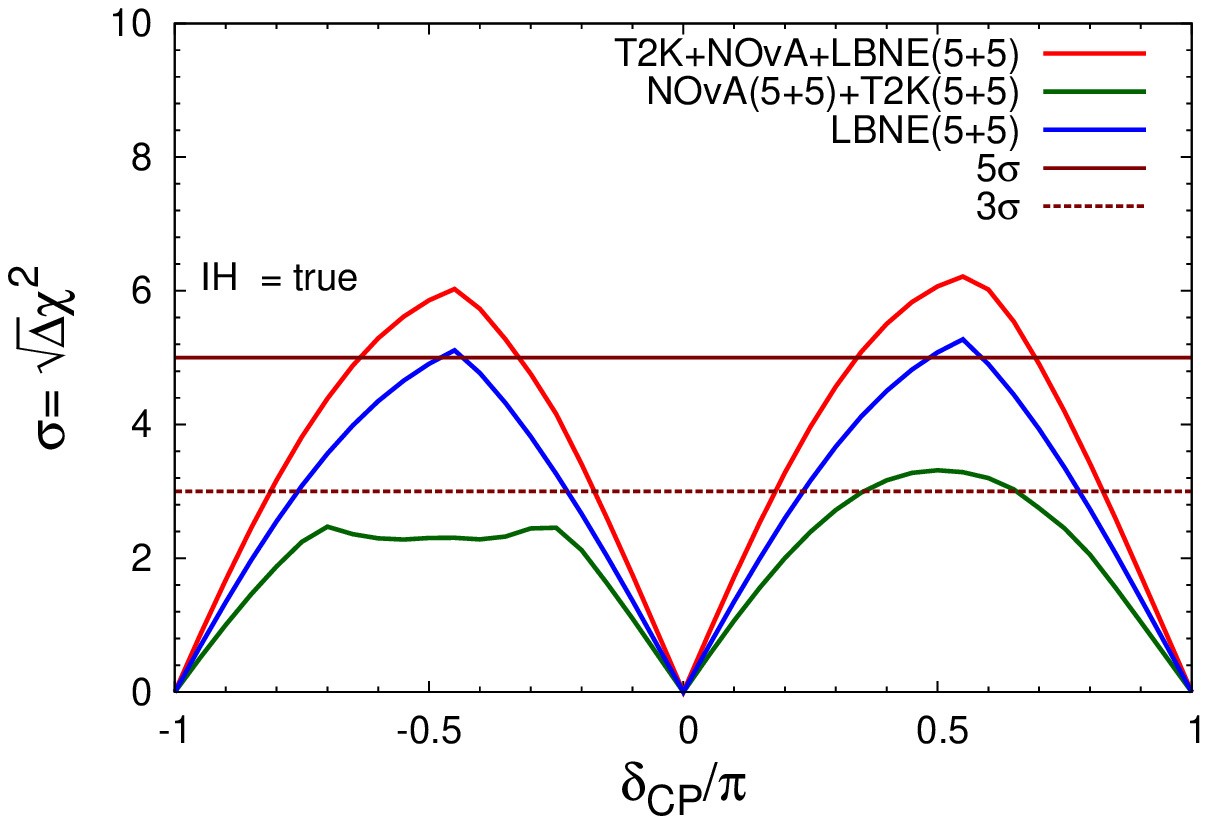}
\caption{Plots for CP Violation sensitivity. Normal hierarchy (Inverted hierarchy) is considered as true hierarchy for a running of (5+5) yrs of T2K, NO$\nu$A, LBNE in $\nu+\bar{\nu}$ mode in the left (right) panel.}
\end{center}
\end{figure}
 From the Fig.~3, we can see that for both T2K and NO$\nu$A
experiments the significance to determine leptonic CP violation phase is almost below
$3 \sigma$  for 5+5 years of run time. The CP violation sensitivity for LBNE experiment is above 
$3 \sigma$ for nearly 40\% of the $\delta_{CP}$ space.
Once we combine all the three experiments T2K+NO$\nu$A+LBNE, we can see that 
almost for about 50\% true values of $\delta_{CP}$ we can measure the leptonic CP violation phase with $3 \sigma$ confidence.

To understand the role of the cross-section uncertainties in the determination of CP violation for  LBNE  experiment
 we consider two  optimistic sets of errors of 10\% and 20\% on the individual cross-sections of $\nu_{\mu}$ and $\nu_{e}$
 in our analysis.
 The bands on the top  panel of Fig. 4 represent the effect due to  10\% uncertainty on the individual cross-sections, whereas the plots 
 in the bottom panel
  show the 20\% cross-section uncertainty effects. Thus, as seen from these figures the CP violation sensitivity affected   
   significantly by the cross-section uncertainties. Furthermore, it should also be noted that the region close to maximal CP violation,
   (i.e.,  $\delta_{CP}=\pi/2$) affected much due to these uncertainties. Also, as generally anticipated, there is an 
   enhancement in these uncertainties with the increase in detector volume as  the possibility of occuring neutrino-nucleus interaction
will be more and hence the uncertainty will be more. 


\begin{figure}[!htb]
\begin{centering}
\begin{tabular}{cc}
\includegraphics[width=0.49\columnwidth]{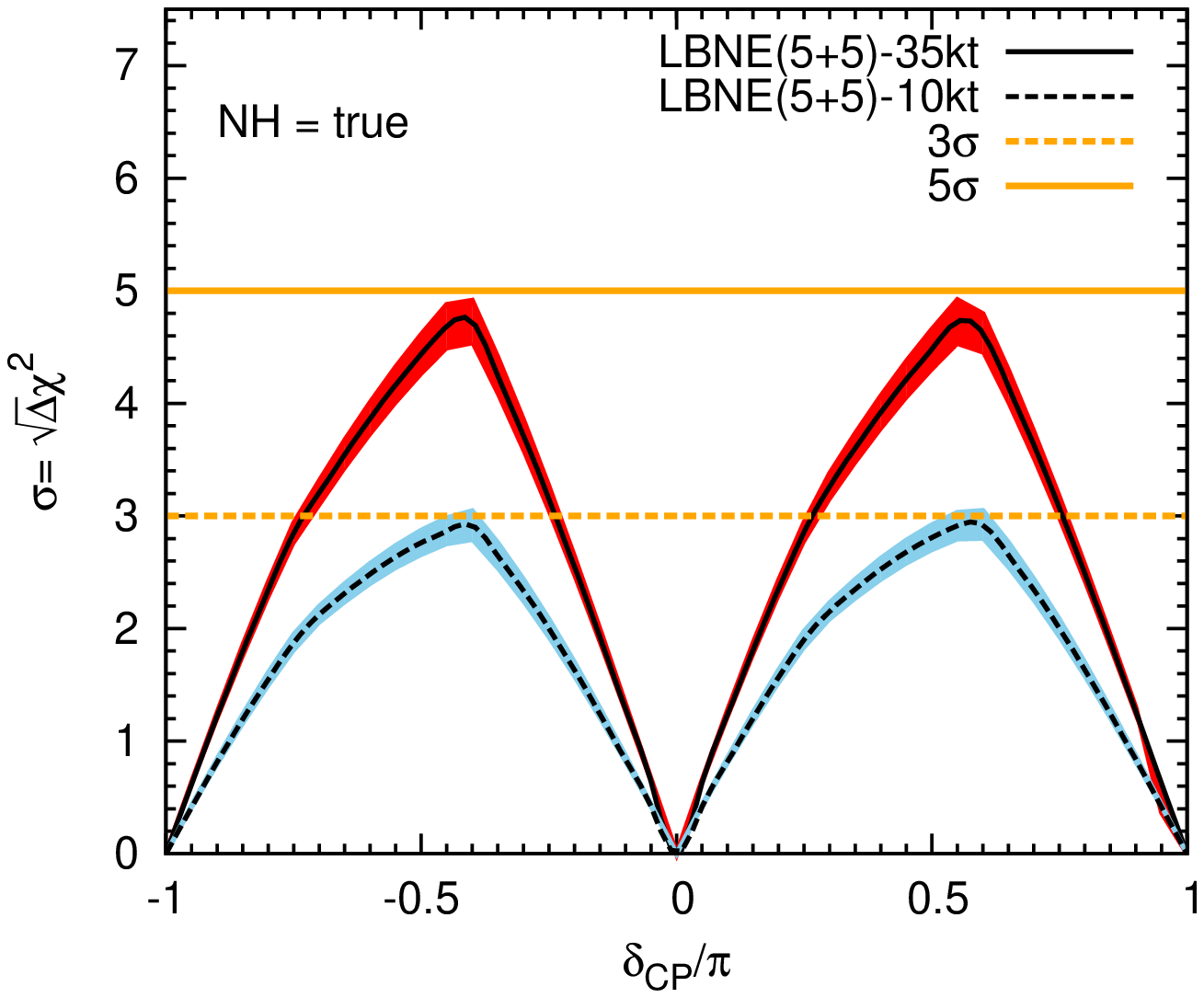}
\includegraphics[width=0.49\columnwidth]{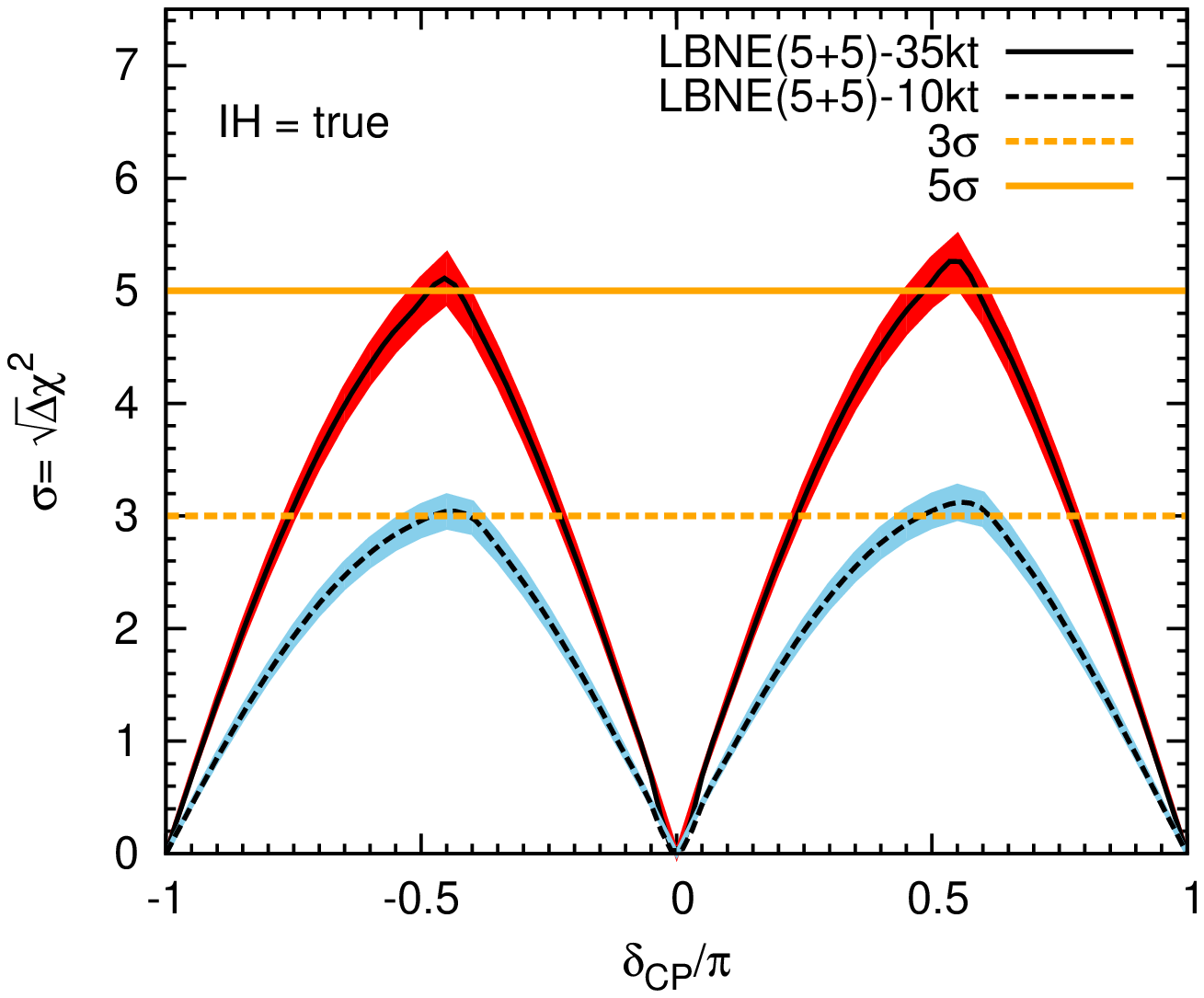}\\
\includegraphics[width=0.5\columnwidth]{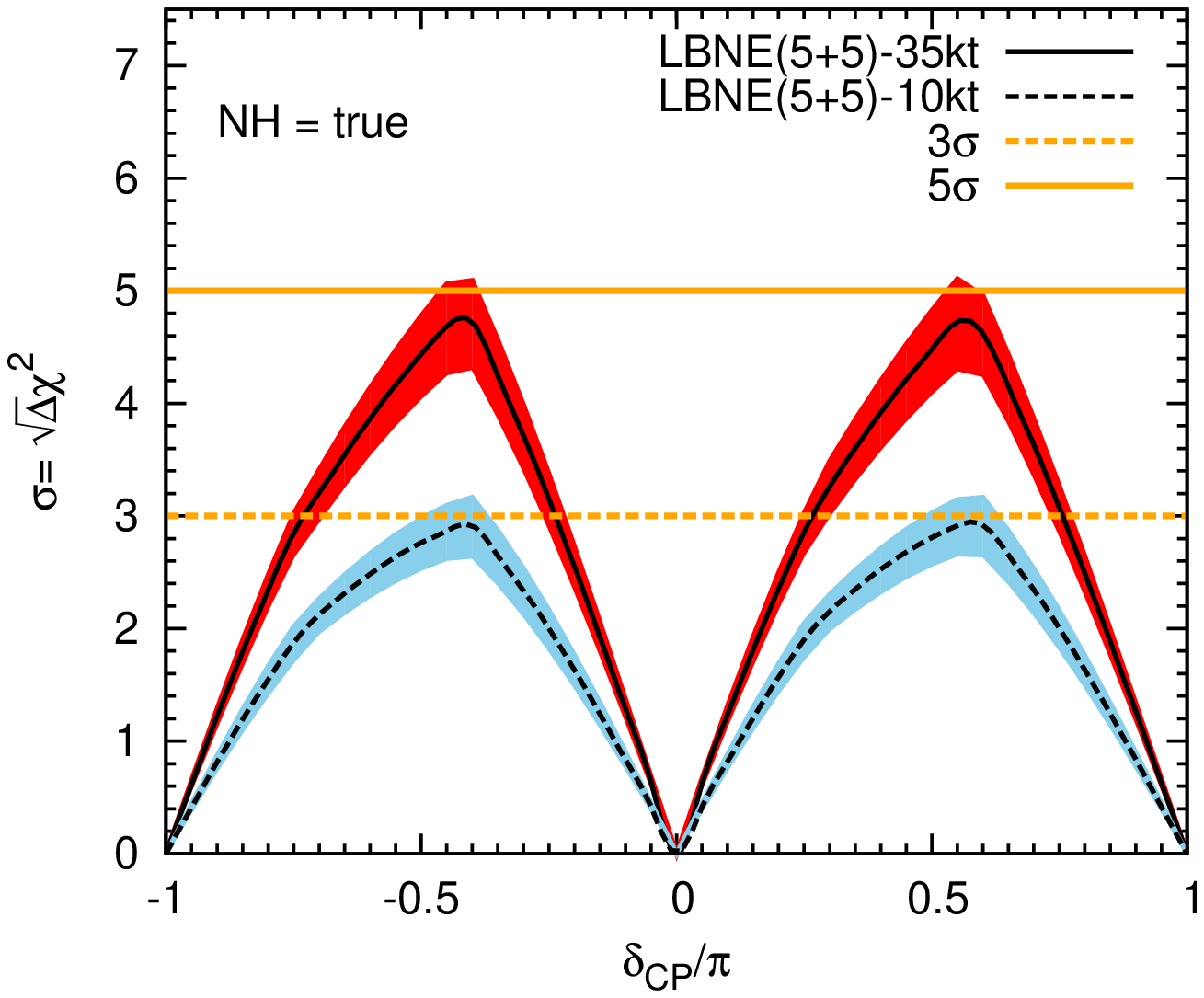}
\includegraphics[width=0.5\columnwidth]{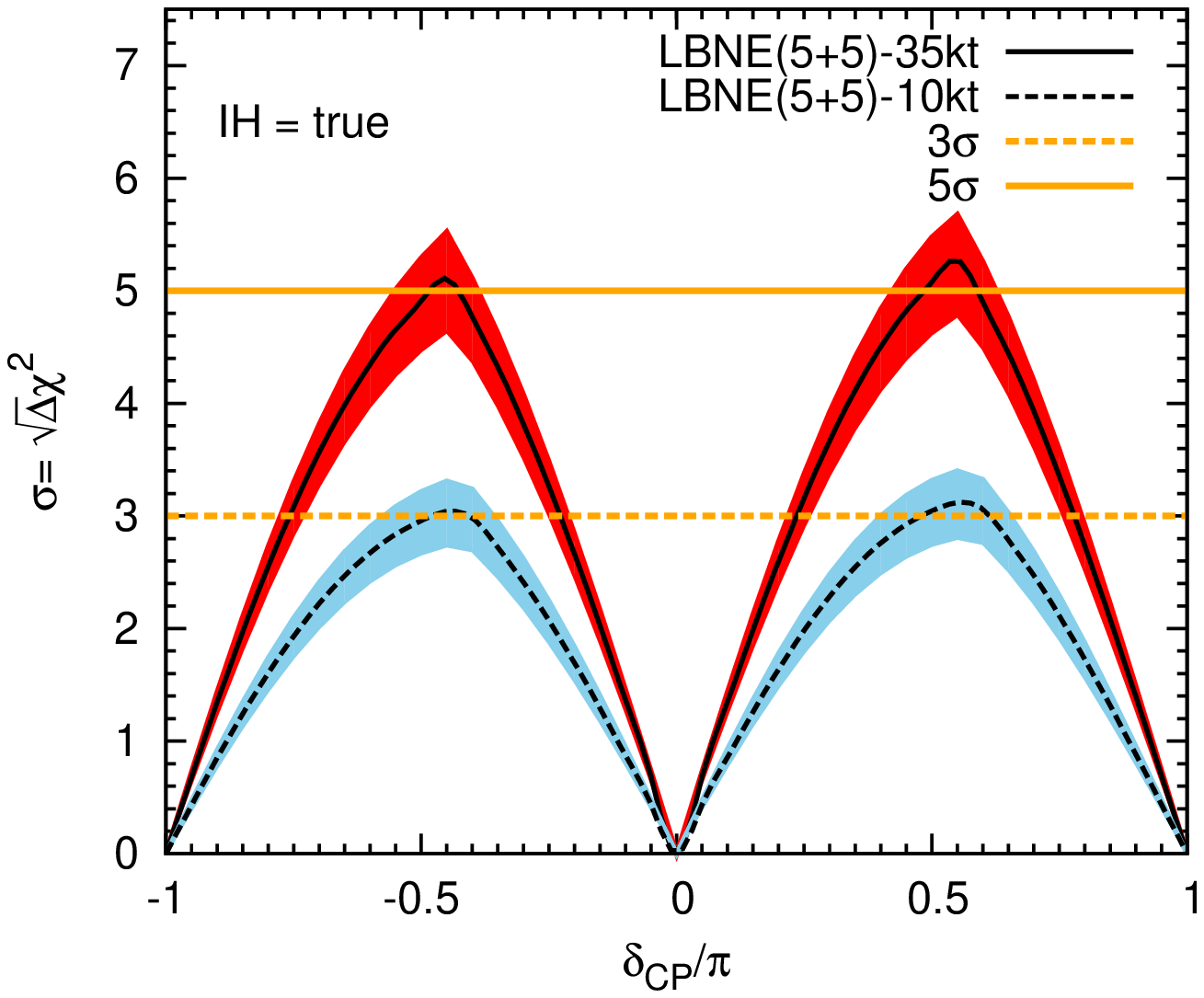}
\end{tabular}
\par\end{centering}
\caption{CP violation sensitivity plots for a representative  10\% (top panel) and  20\%  (bottom panel) errors on the individual cross-sections of $\nu_{\mu}$ and $\nu_{e}$ for 5+5 years run of LBNE experiment.  }
\end{figure}
\section{Determination of Mass Hierarchy}
Long-baseline experiments such as NO$\nu$A, T2K and LBNE primarily use the $\nu_\mu \to \nu_e$  and the corresponding anti-neutrino oscillation
channels to determine the neutrino mass hierarchy (MH). Using the approximate perturbative
formula for the probability $P_{\mu e}$, it can be seen that there is a hierarchy-$\delta_{CP}$ degeneracy
as discussed in section 2. As a result, the hierarchy sensitivity of these experiments is a strong function of
the value of the CP violating phase $\delta_{CP}$.
The value of $\Delta \chi^2$ for the mass hierarchy study can be obtained using the relation
 \begin{equation}
\Delta \chi^{2}=\left|\chi^{2}_{{MH}^{test}}-\chi^{2}_{{MH}^{true}}\right|\;.\label{mh} 
\end{equation}
We have used Eq. (\ref{mh}) to obtain the mass hierarchy significance and  considered the following two cases.
In the former, we consider true hierarchy to be normal hierarchy (NH) and obtain the values of
$\Delta \chi^{2}$ by assuming inverted hierarchy as test hierarchy.   In the later, we consider true hierarchy to be inverted hierarchy (IH) and
assume normal hierarchy to be the test hierarchy while obtaining the $\Delta \chi^{2}$ value. The true
values for the oscillation parameters are used from Eq. (\ref{true-values}) and 
 the test values of $\Delta m_{atm}^{2}$, $\sin^2 \theta_{23}$, and $\sin^2 2 \theta_{13}$ are
marginalised  over their 3$\sigma$ ranges in both the cases. 
We also added prior for $\sin^22\theta_{13}$ with $\sigma(\sin^22\theta_{13}) =0.01$. In
Fig.~5, we present the resultant significance plots.
The 3$\sigma$ and 5$\sigma$ lines correspond to $\Delta \chi^{2}$ = 9 and 25 (for 1 degree of freedom)  which indicate approximately
$99.73\%$ and $99.99\%$ probability of determining the correct mass hierarchy respectively.
The values of $\delta_{CP}$ for which the curve is above 3$\sigma$ (5$\sigma$) are the values
for which hierarchy can be determined with  $99.73\%$ $(99.99\%)$ confidence level.


\begin{figure}[!htb]
\begin{center}
\includegraphics[width=0.49\columnwidth]{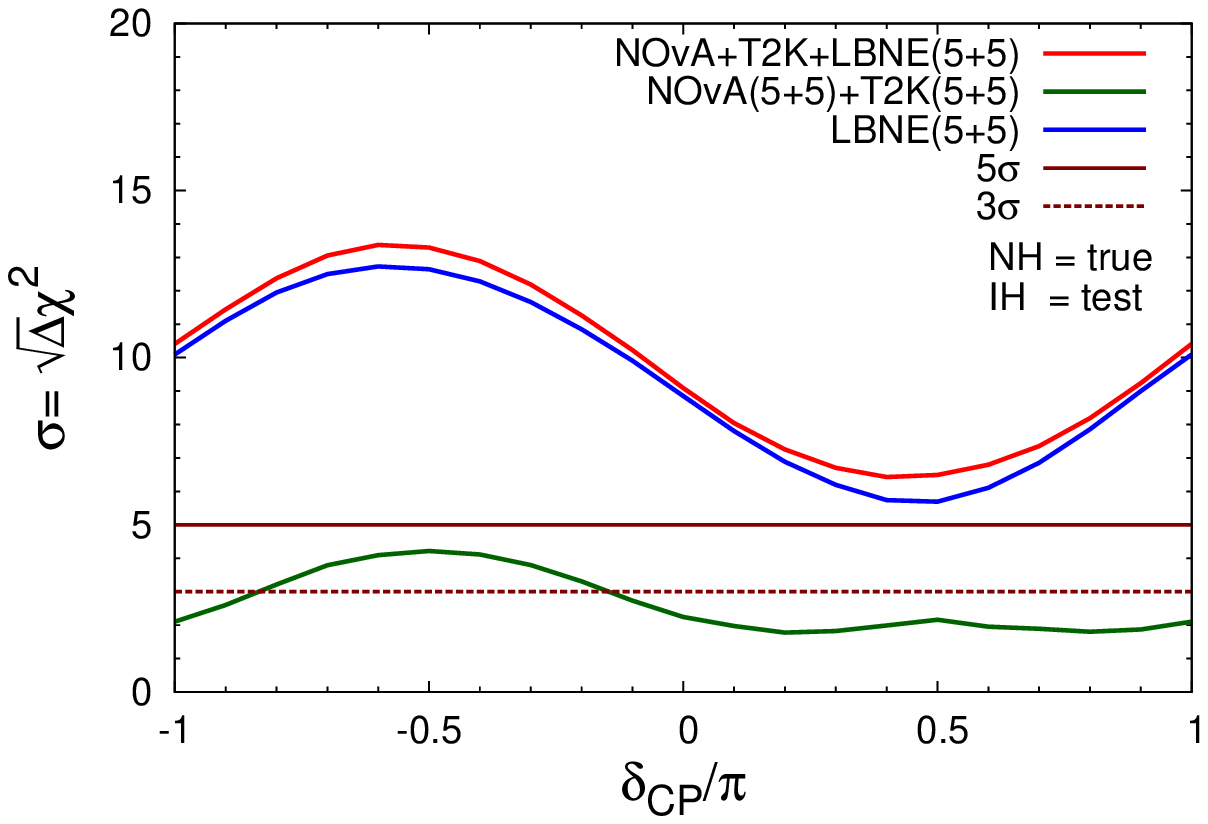}
\includegraphics[width=0.5\columnwidth]{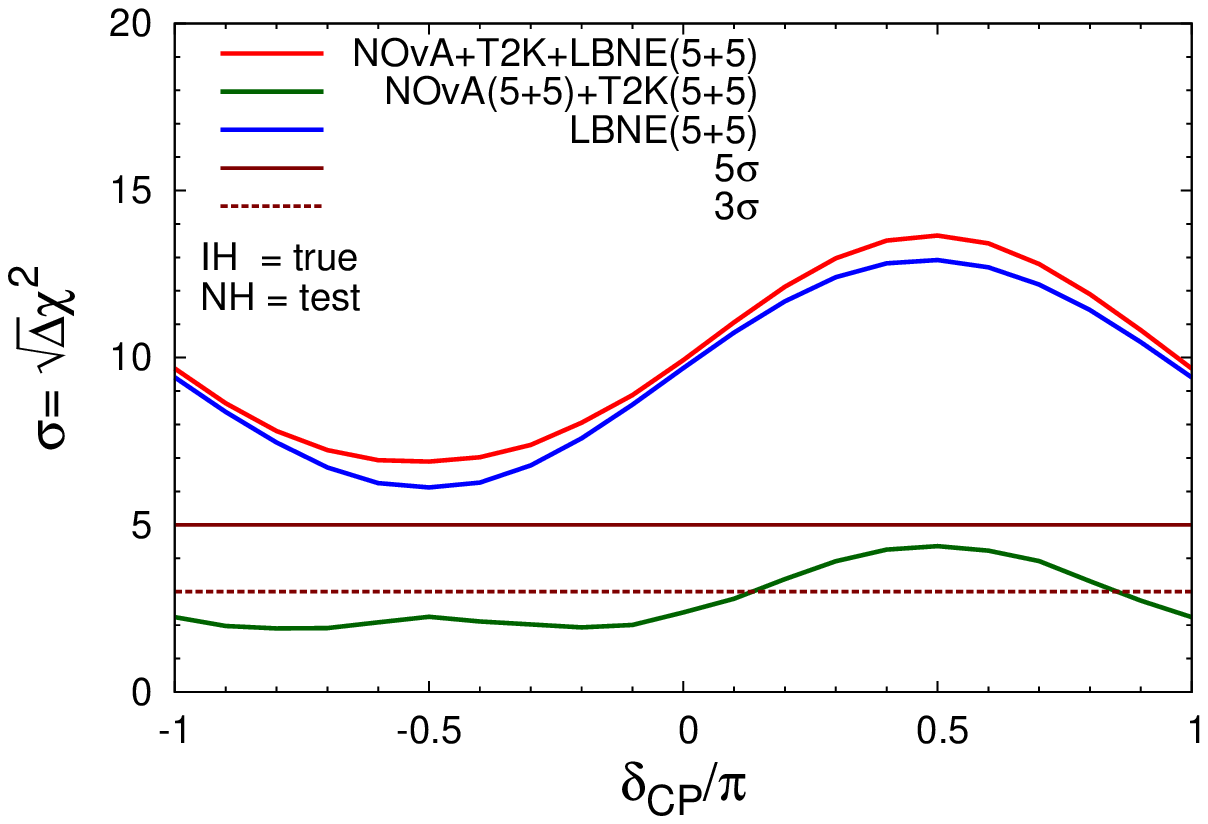}
\end{center}
\par\caption{Mass hierarchy significance as a function of true $\delta_{CP}$. Normal hierarchy (Inverted hierarchy) is considered as true hierarchy and inverted (normal) is taken as test hierarchy for a running of (5+5) yrs of T2K, NO$\nu$A, LBNE in $\nu+\bar{\nu}$ mode in the left (right) panel.}
\end{figure}

From   Fig.~5, we can see that both T2K and NO$\nu$A experiments have significance to mass hierarchy less than $5\sigma$ when run for 10 years in normal 
and inverted hierarchies. LBNE when run for $(5+5)$ years has more than $5\sigma$ significance to measure the hierarchy for all the values of $\delta_{CP}$. 
 When we combine the results from all the three experiments T2K+NO$\nu$A+LBNE run for
10 years each, there will only be marginal improvement on the mass hiearachy sensitivity, over the LBNE result.

\section{Octant Sensitivity of $\theta_{23}$}
In this section we would like to analyze the capabilities of current long baseline neutrino oscillation
experiments and LBNE to measure octant sensitivity.
As in the case of mass hierarchy  determination, adding information from various experiments enhances the
sensitivity. However, it is the precise knowledge of the value of $\theta_{13}$ that plays a crucial role in
determining the octant correctly.

Using atmospheric neutrino oscillations Super-Kamiokande has measured the value of $\sin^2(2\theta_{23})$ to be $ >0.95$ at
$90\% $  confidence level \cite{sk}.
This corresponds to a value of $\theta_{23}$ around  $45^{\circ}$ leaving an ambiguity of whether the value of
$\theta_{23}$ is less than $45^\circ$ i.e., in the lower octant or in the upper octant where
$\theta_{23}$ is greater than  $45^\circ$.

For the T2K and NO$\nu$A which are off-axis experiments with baselines 295 and 810 km, the beam energies peak at 0.6 and 2 GeV respectively.
However, LBNE ($L$=1300 km) which is an on-axis experiment has a broad band beam (0.5 GeV to 8 GeV) covering first and second oscillation maxima,
with minimal high energy tail above $\sim  5$ GeV. For these values of baseline lengths,
 the earth matter density varies in the range 2.3-2.8 g/cc, and the corresponding matter resonance energies are above 10 GeV.
Hence the neutrino energies of these experiments lie well below the matter resonances and the oscillation probabilities will have
only very small sub-leading matter effects. Thus, the expressions for the relevant oscillation probabilities (in vacuum) can be obtained by
assuming  one-mass scale
dominance (OMSD) approximation \cite{prob1} as
\begin{equation}
{\mathrm{P^{v}_{\mu e}}} =
{\mathrm{
\sin^2 \theta_{23} ~ \sin^2 2 \theta_{13} ~ \sin^2\left[1.27  \frac{\Delta m_{31}^2L}{E} \right]
}}\;.
\label{eq:pemu}
\end{equation}
This equation depends on the combination of mixing angles $\sin^2 \theta_{23} \sin^2 2\theta_{13}$.
Thus, there exists a correlation between $\sin^2 \theta_{23}$  and $\sin^2 2 \theta_{13}$ which
implies that for different values of $\theta_{13}$ there can be values of $\theta_{23}$ in
opposite octants which give same value of oscillation probability.

The disappearance probability of muon neutrino beam is given by
\begin{equation}
{\mathrm{P^{v}_{\mu \mu}}} =
{\mathrm{
1 -  \sin^2 2 \theta_{23} \sin^2\left[1.27 \frac{\Delta m_{31}^2L}{E} \right] + 4 \sin^2 \theta_{13} 
\sin^2 \theta_{23}  \cos 2 \theta_{23}  \sin^2\left[1.27 \frac{\Delta m_{31}^2L}{E} \right].
}}
\label{eq:pmmu}
\end{equation}
The leading order term in the above equation has its entire dependency on $\sin^2 2 \theta_{23}$ giving rise to intrinsic octant degeneracy.
The oscillation probability expressions (9) and (10) mentioned above  are  only  given for  illustration of octant degeneracy  and  we have considered the  
full   formula of oscillation probability  in  our  calculations.

We first look into the bi-probability plots for LBNE experiment with (5+5) years of run, to estimate its capabilities in determining
 mass hierarchy and resolving octant degeneracy. The left panel of Fig.~6 shows $\nu$ appearance events vs. $\bar{\nu}$ appearance events 
for all combinations of octant-hierarchy. Here the red curves are obtained by considering normal hierarchy mass ordering, 
LO (HO) i.e.,  $\sin^2\theta_{23}=$0.41 (0.59) and the blue curves are obtained by considering inverted hierarchy mass ordering 
for LO (HO). We plot these ellipses by obtaining the event spectra for (5+5) yrs of runs in $\nu$ and $\bar{\nu}$ mode for 
LBNE experiment for all values of $\delta_{CP}$.

 For the analysis  of octant determination of $\theta_{23}$, we have used GLoBES to evaluate $\Delta \chi^2$. In our simulation, 
we have kept true values of oscillation parameters as $\sin^2\theta_{12}=$0.32, $\sin^2 2\theta_{13}=$0.1, 
$\Delta m_{21}^2= 7.6 \times 10^{-5}~{\rm eV}^2$, $\delta_{CP}=0$ and $\Delta m_{atm}^2 =2.4 \times 10^{-3}~{\rm eV}^2$ (NH). 
Furthermore, we have done marginalization  over test values in the following ranges: for $\sin^2 2\theta_{13}$ and $\Delta m_{atm}^2$ 
in their $3\sigma$ ranges, for $\delta_{CP}$ in its full range and for $\sin^2\theta_{23}$ in LO for true higher octant and HO for 
true lower octant. We have also added priors for $\sin^2 2\theta_{13}$ and $\sin^2\theta_{23}$ with $\sigma(\sin^2 2\theta_{13})=0.01$ and  
$\sigma(\sin^2\theta_{23}) =0.05$.

 In the right panel of Fig. 6, we illustrate the ability of NO$\nu$A+T2K+LBNE to determine the octant as a function of the
true value of $\theta_{23}$. We see that with LBNE and NO$\nu$A+T2K+LBNE, the octant
can be determined at \textgreater$5 \sigma$ C.L. when $\sin^2\theta_{23} \approx 0.41$. 
 For  values  of  $\theta_{23}$  above 40$^\circ$, (i.e, for  $0.5 \geq \sin^2 \theta_{23} > 0.41$),  
the  sensitivity  to  solve  the octant  degeneracy  is  reduced,  well  below  5$\sigma$.

\begin{figure}[htb]
\includegraphics[width=8cm,height=7cm, clip]{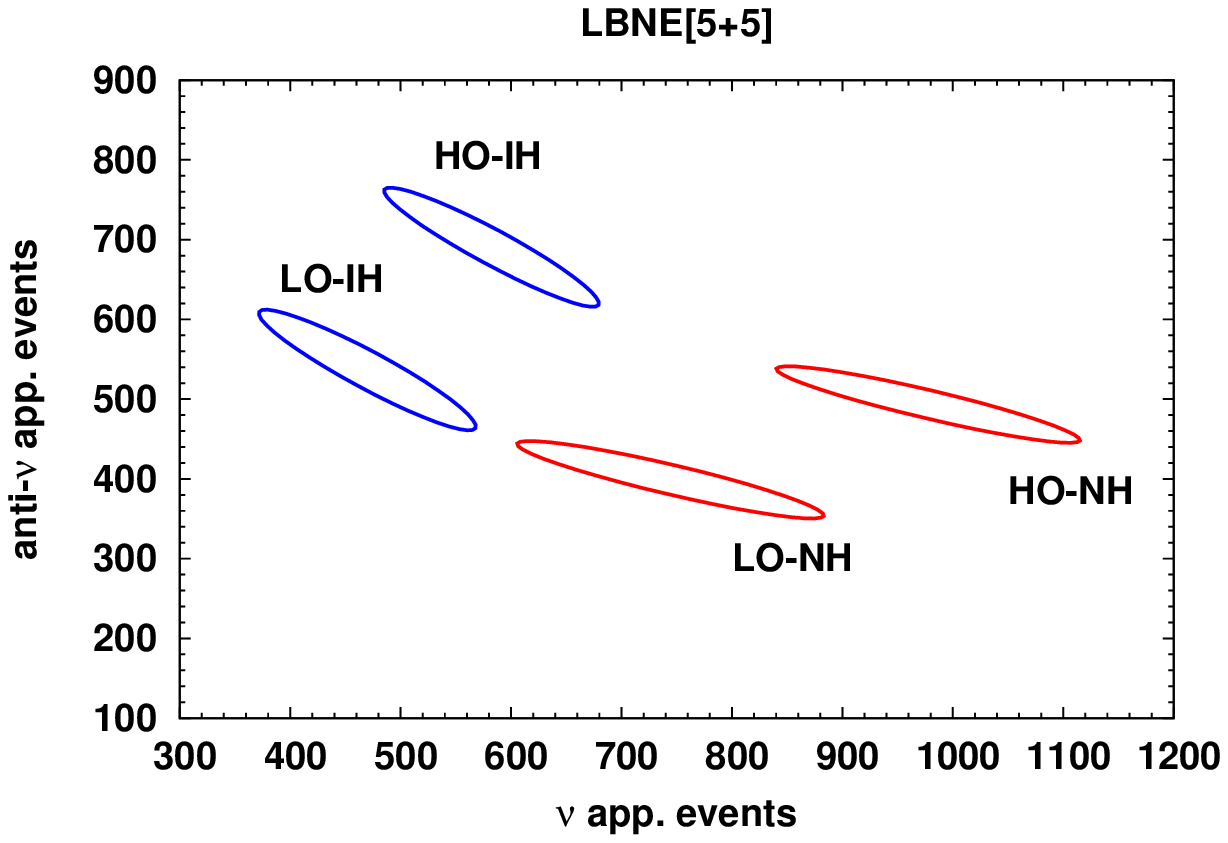}
\hspace{0.2 cm}
\includegraphics[width=8cm,height=7cm, clip]{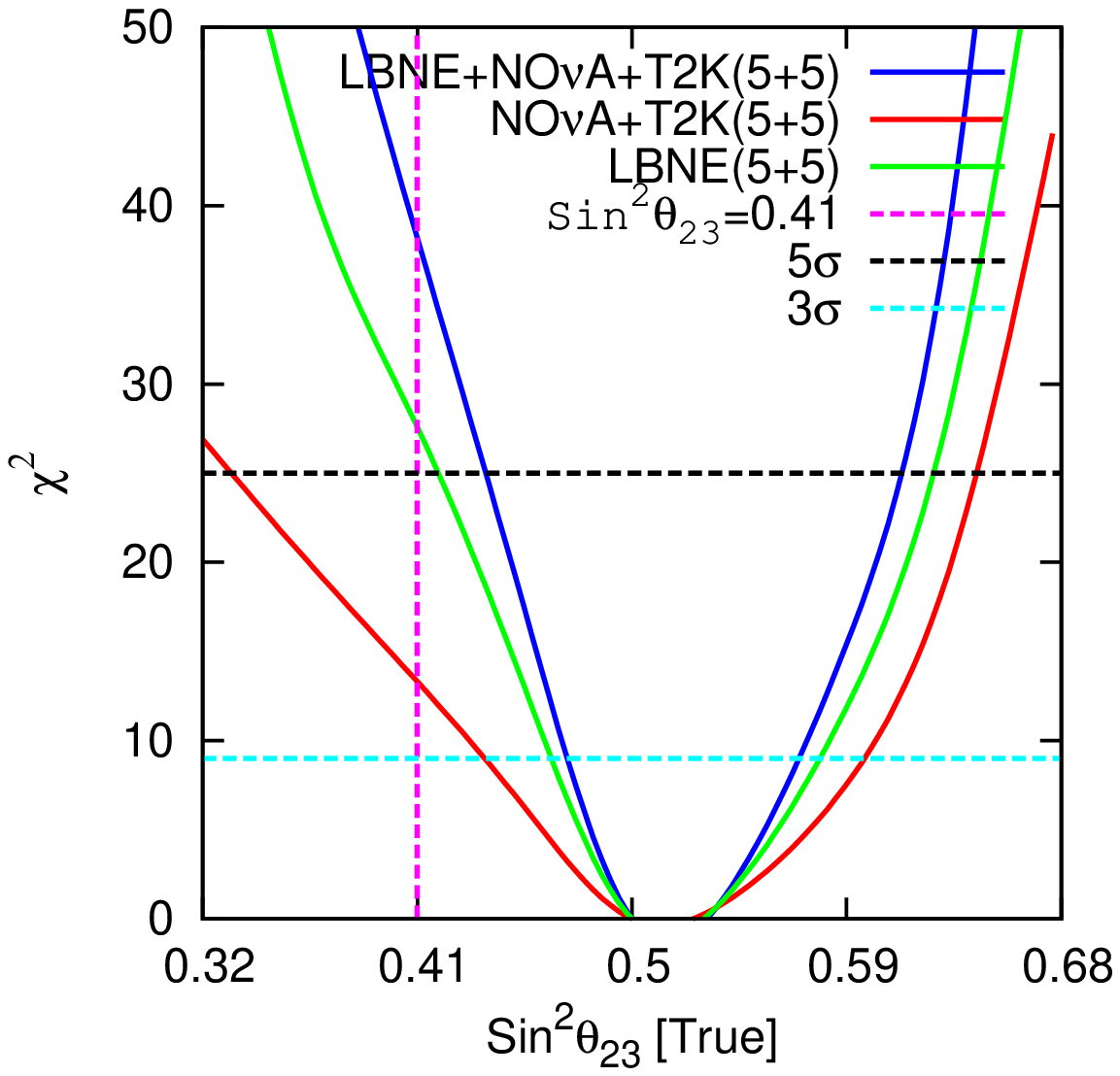}
\caption{In the left panel: Neutrino and anti-neutrino appearance events in LBNE (5+5)yrs for $\nu_{\mu} \rightarrow \nu_{e}$ versus $\bar{\nu}_{\mu} \rightarrow \bar{\nu}_{e}$ for all combinations of mass hierarchy-octant. In the right panel: Octant sensitivity for a running of (5+5)yrs for T2K, NO$\nu$A, LBNE in $\nu+\bar{\nu}$ mode}
\label{CPV-4}
\end{figure}

\subsection{Allowed regions in test $\delta_{CP}$ and test $\sin^2\theta_{23}$ plane}
Next, we would like to study the correlation between $\delta_{CP}$ and  $\sin^2\theta_{23}$ for different combinations of true 
hierarchy and true octant. For our study, we simulate data for 5+5 yrs of run of LBNE, T2K+NO$\nu$A and LBNE+T2K+NO$\nu$A. 
We have taken true  $\delta_{CP}=0$, assumed true hierarchy as NH  and true octant as LO/HO. We have varied the test values  of 
$\sin^2\theta_{23}$  in the range [0.32:0.68] and that of $\delta_{CP}$ in its full range [$-\pi : \pi$]. 
We have done marginalization over $\sin^22\theta_{13}$ and  $\Delta m_{31}^2$ and added prior for $\sin^22\theta_{13}$ with 
$\sigma(\sin^22\theta_{13}) =0.01$. Finally, we calculated the  minimum $\chi^2$  over all these test parameter combinations.
The obtained result is then studied as a function of $\delta_{CP}$(test) and $\sin^2\theta_{23}$(test).
We have plotted the 2$\sigma$ contour in the space spanned by test values of $\delta_{CP}$ and $\sin^2\theta_{23}$  for LBNE, T2K+NO$\nu$A and LBNE+T2K+NO$\nu$A.


\begin{figure}[!h]
\begin{centering}
\begin{tabular}{cc}

\includegraphics[width=0.49\columnwidth]{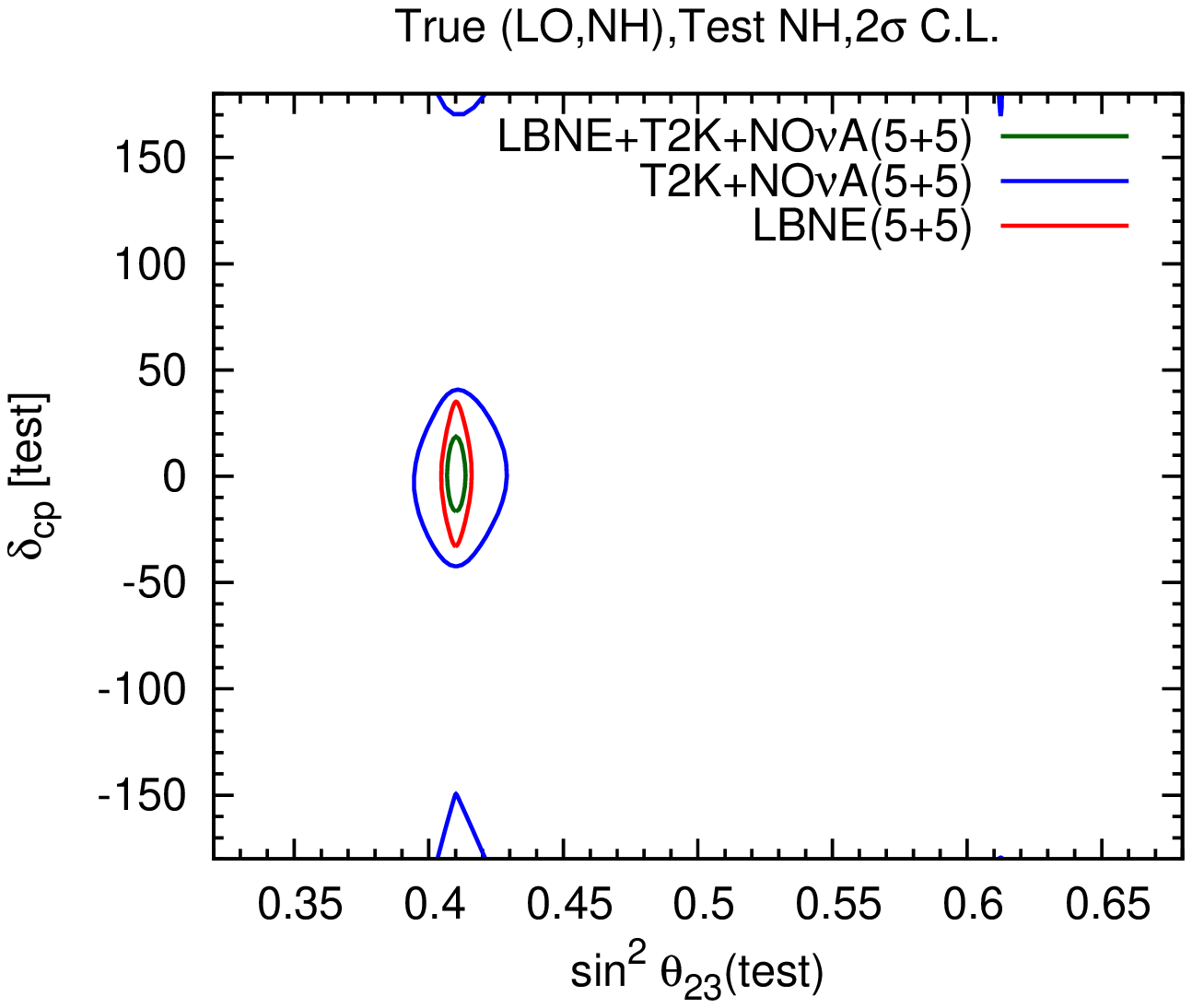}
\includegraphics[width=0.5\columnwidth]{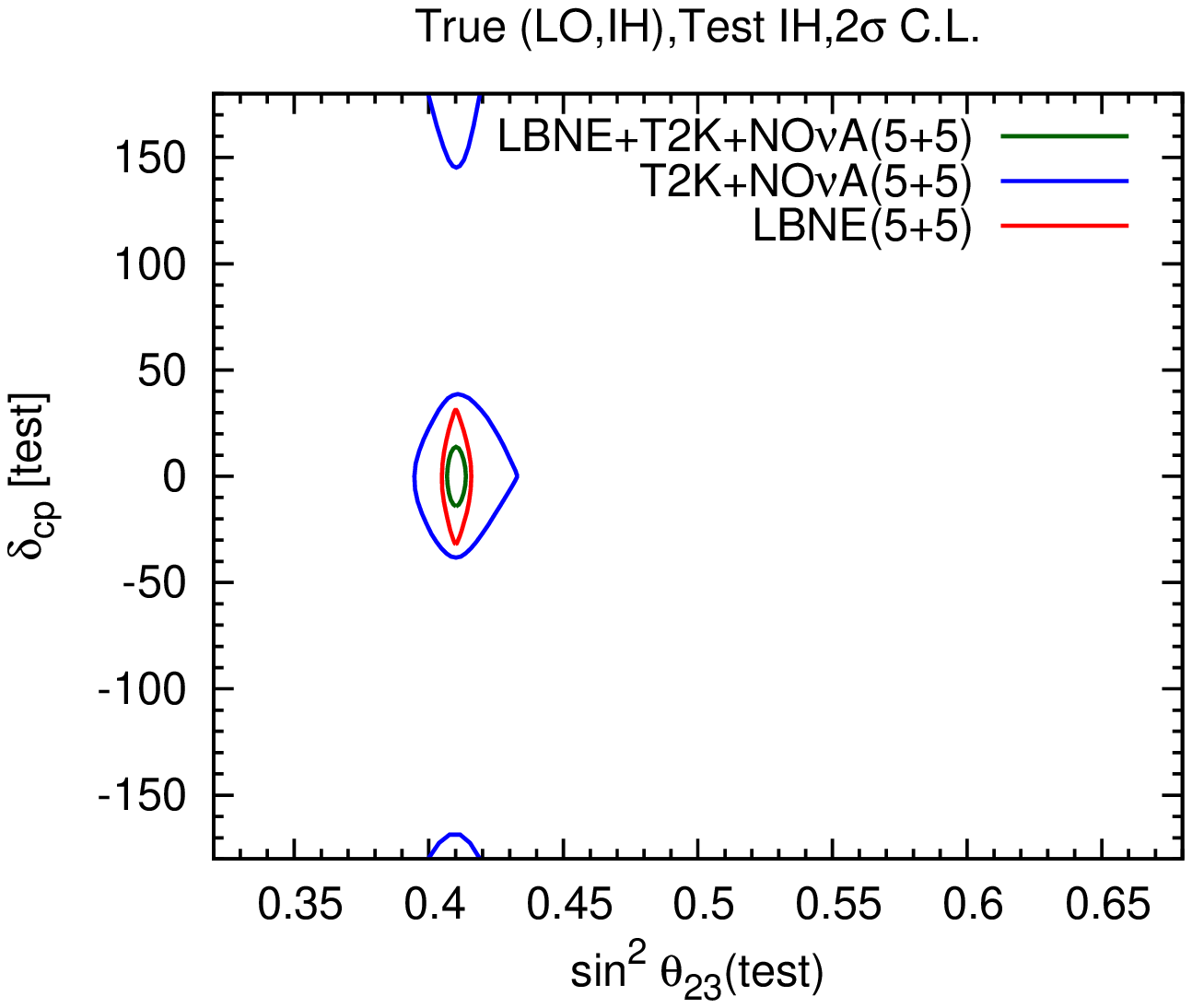}\\
\includegraphics[width=0.49\columnwidth]{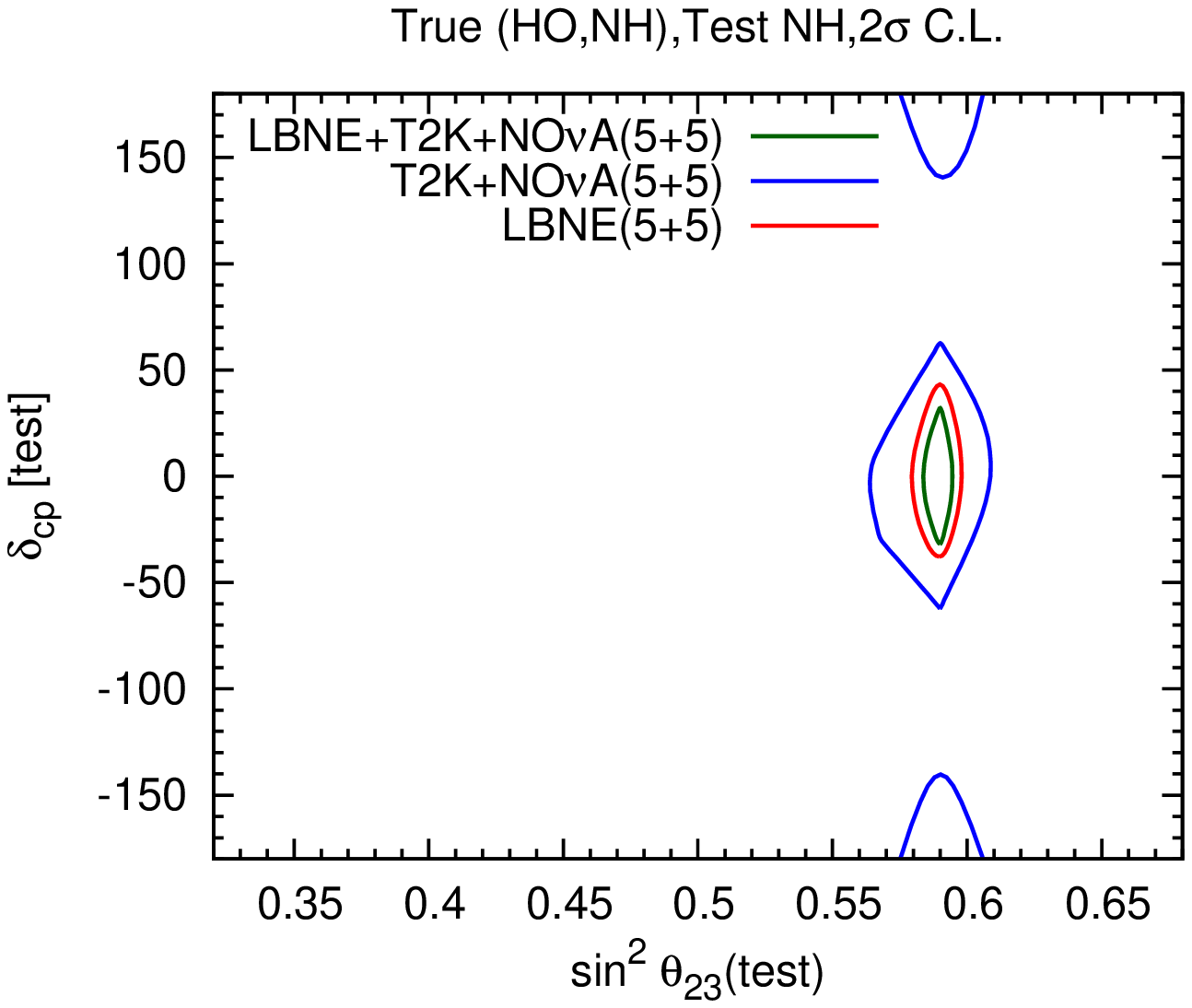}
\includegraphics[width=0.5\columnwidth]{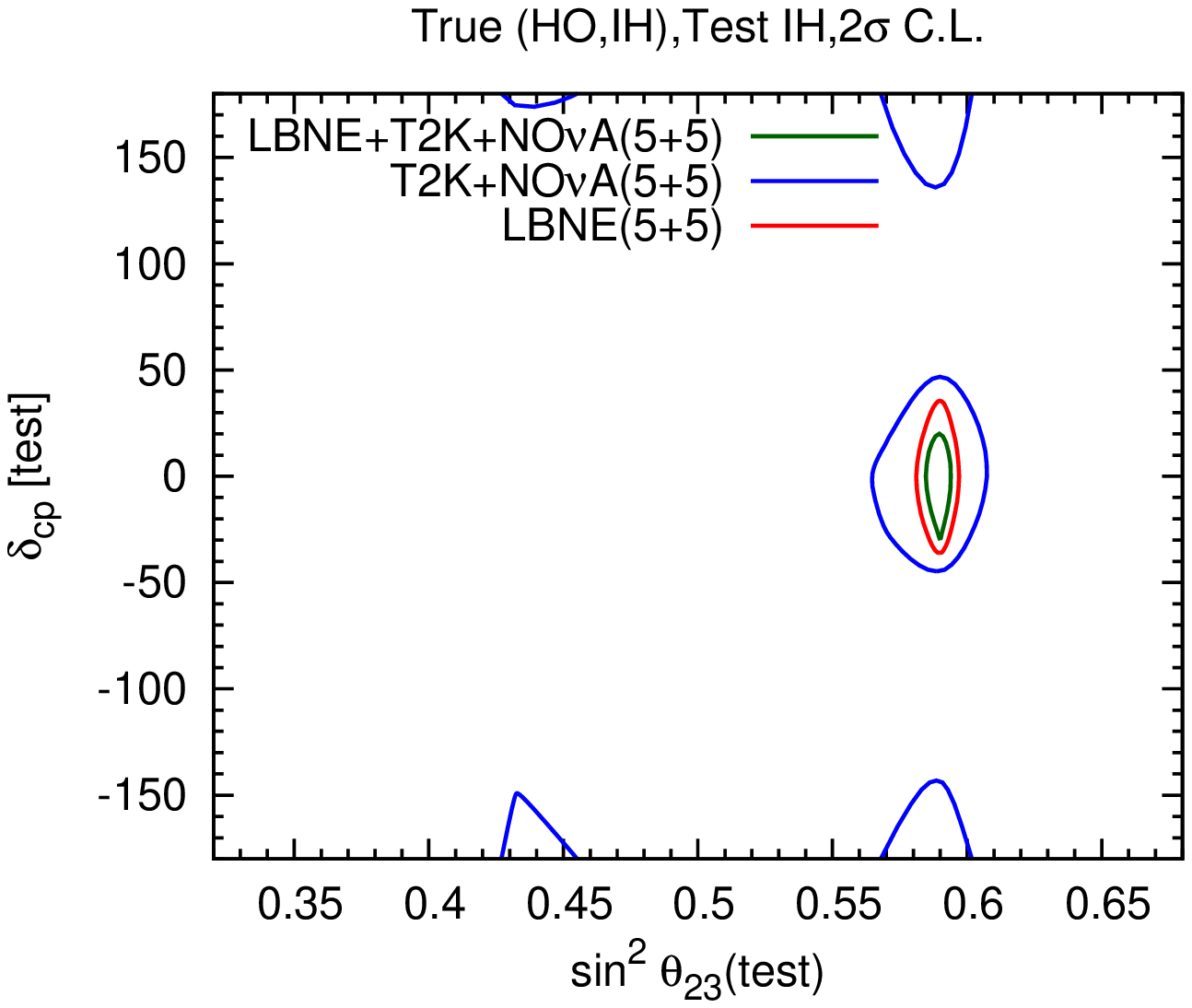}

\end{tabular}
\end{centering}
\par\caption{Allowed regions in  $\delta_{CP}$ and  $\sin^2\theta_{23}$ plane at $2\sigma$ C.L. for true $\delta_{CP}$=0 with  combinations of  5+5 yrs of run of LBNE, T2K+NO$\nu$A and LBNE+T2K+NO$\nu$A. The top (bottom) plot corresponds to LO (HO) with known hierarchy as NH (IH) in left (right) plot.}

\end{figure}

Fig. 7 shows the allowed regions in  $\delta_{CP}$ and  $\sin^2\theta_{23}$ plane at $2\sigma$ C.L. for true $\delta_{CP}$=0 with  
combinations of  5+5 yrs of run of LBNE, T2K+NO$\nu$A and T2K+NO$\nu$A+LBNE. The top (bottom) plot corresponds to LO (HO) with known 
hierarchy as NH (IH) in left (right) plot.  Thus, from these plots one can tightly constrain the allowed value of $\delta_{CP}$ as 
well as discriminate the wrong octant of $\theta_{23}$.

\section{Correlation between $\delta_{CP}$ and $\theta_{13}/\theta_{23}$}

In this section we present the correlations between the CP violating phase
$\delta_{CP}$ and the mixing angles $\theta_{13}/\theta_{23}$.

The correlation between $\delta_{CP}$ and $\theta_{13}$
 is obtained by projecting $\chi^{2}$ onto the two-dimensional plane of $\delta_{CP}$ and $\theta_{13}$
by spanning over test values of $\delta_{CP}=[-\pi:\pi] $ and $\sin^2(2\theta_{13})\in [0.07,0.13]$.
Here also we have marginalized over $\theta_{23}$ and $\Delta m^{2}_{31}$. We have obtained 1$\sigma$,
2$\sigma$ and 3$\sigma $ contours by considering three true values for $\delta_{CP}=0,-\pi/2, +\pi/2$.
We have set $10 \%$ error on each of the solar parameters and a $5 \%$ error for the matter
density and assumed the hierarchy to be normal. The result is presented on the left panel of Fig. 8,
where the  blue/green/red curves  correspond to 1/2/3$\sigma$ measurement contours for a total (5+5) yrs
running of T2K+NO$\nu$A+LBNE.

\begin{figure}[htb]
\includegraphics[width=8cm,height=6.75cm, clip]{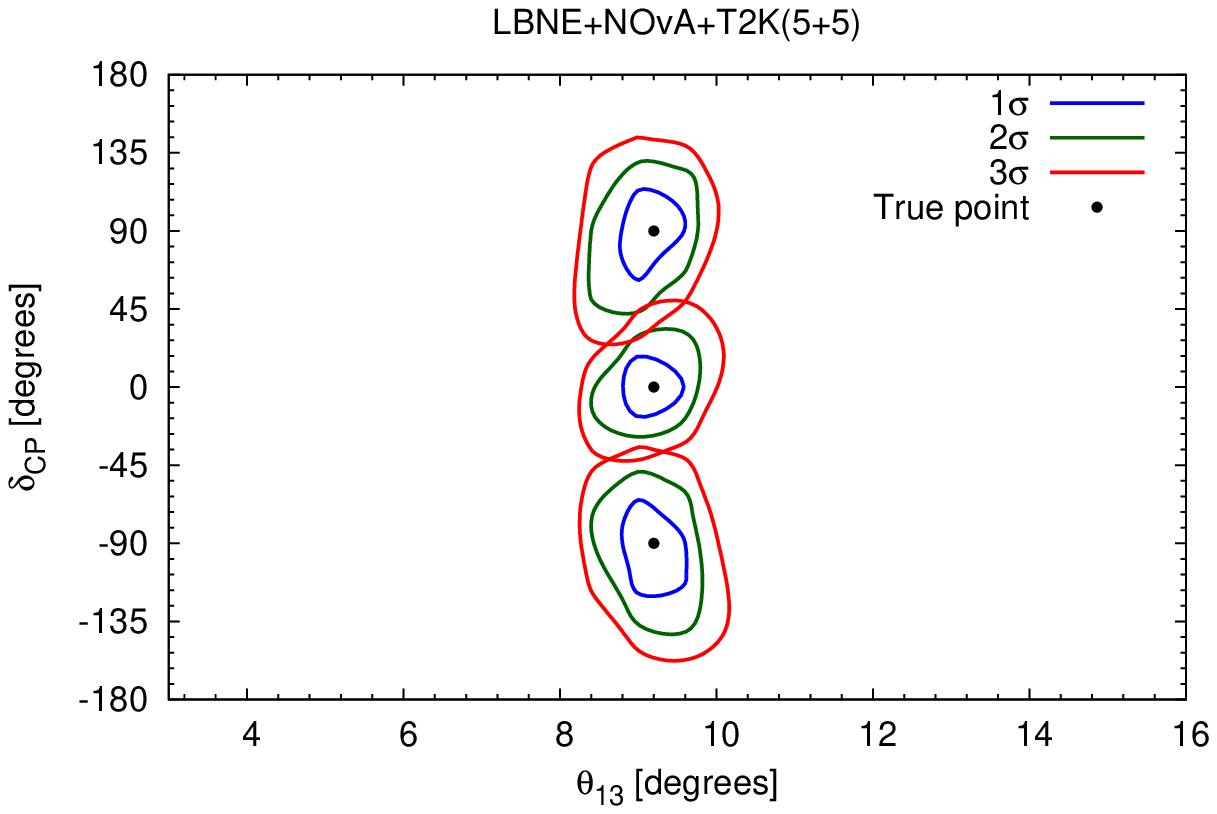}
\hspace{0.2 cm}
\includegraphics[width=8cm,height=6.75cm, clip]{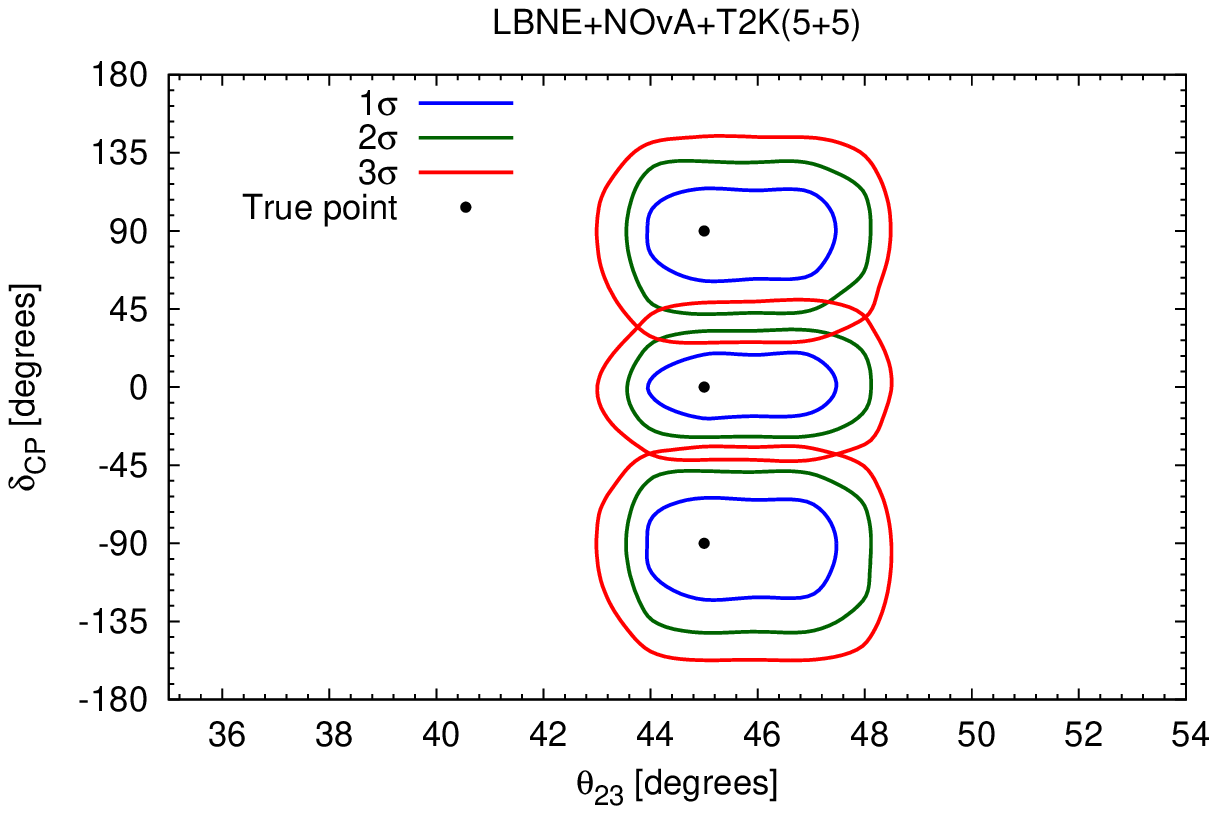}
\caption{Confidence region in $\theta_{13}-\delta_{CP}$ plane (left panel) and  $\theta_{23}-\delta_{CP}$ plane (right panel)
for a running of (5+5) yrs for T2K+NO$\nu$A+ LBNE (right panel) in $\nu+\bar{\nu}$ mode.}
\label{CPV-4}
\end{figure}

The analogous plot  between $\delta_{CP}$ and $\theta_{23}$  has been obtained  following
similar procedure and the corresponding result is shown in the right panel of Fig. 8.
The allowed region is very tightly constrained which indicates that with the combined  10 years of LBNE+T2K+NO$\nu$A data, it is
possible to constrain the value of $\delta_{CP}$. The overlap in the 3$\sigma$ contours in both the plots can be accounted by the fact that 
we are considering three different true values for $\delta_{CP}=0,-\pi/2, +\pi/2$.

\section{Summary and Conclusion}

In this paper we have explored the possibility of determining the mass hierarchy, octant of the atmospheric
mixing angle $\theta_{23}$ and the CP violation discovery potential in the current generation and upcoming
long baseline experiments T2K, NO$\nu$A and LBNE and our findings are summarized below.

$\bullet $ For long-baseline experiments, it is well known that the measurement of the mass
hierarchy is easier than a measurement of $\delta_{CP}$ because matter effects enhance the separation
between the oscillation spectra, and hence the event rates between normal and inverted hierarchies.
The determination of  mass hierarchy is defined as the ability to exclude any degenerate solution
for the wrong (fit) hierarchy at a given confidence level. From our analysis, we find that if we combine the results
from all the three experiments for (5$\nu+ 5 \bar{\nu}$) years of run we can determine the mass hierarchy of neutrinos
above $5\sigma$.

$\bullet $ The octant sensitivity of $\theta_{23}$ also increases noticeably if we combine the results of the three
experiments T2K, NO$\nu$A and LBNE.  However, for  values  of  $\theta_{23}$  above
40$^\circ$,  the  sensitivity  to  solve  the  degeneracy  is  reduced,  well  below  5$\sigma$.

$\bullet$ The CP violation discovery potential in the long baseline experiments is also quite promising.
A discovery of CP violation, if it exists, basically means being able to exclude the
CP-conserving values i.e., $\delta_{CP}= 0^\circ $ or $180^\circ$ at a given confidence level.
From our analysis, we found that it is possible to measure the CP violation phase above $3 \sigma$ C.L. for
about $50 \%$ of the true $\delta_{CP}$ range if we combine the data from all three experiments. Furthermore,
it should also be noted that the CP violation measurement becomes very difficult for the $\delta_{CP}$
values which are closer to
$0^\circ$ or $180^\circ$. Therefore, whilst it is possible to
discover the mass hierarchy for all possible values of $\delta_{CP}$,
the same is not true for CP violation.

$\bullet$ The cross-section uncertainties play a crucial role in determining the CP violation sensitivity.
These uncertainties affect significantly the region of maximal-CP violation.

$\bullet$ From the correlation plots between $\delta_{CP}$ and $\sin^2 \theta_{23}$ as well as from
$\delta_{CP}$ and $\theta_{13}/\theta_{23}$ (Figs. 7 and 8), one can see that $\delta_{CP} $ is severely
constrained implying a definitive measurement on $\delta_{CP}$ could be possible with 10 years of LBNE data taking.

In conclusion, we find that combining the data of $(5 \nu+ 5 \bar \nu)$ years of running T2K,
NO$\nu$A and LBNE will help us to resolve most of the ambiguities associated with the neutrino
sector.

\begin{acknowledgments}
KND and SC would like to thank University Grants Commission for financial support. The work of RM was
supported in part by CSIR, Government of India through Grant No. 03(1190)/11/EMR-II. We would like to thank
Drs. S. K. Agarwalla and S. K. Raut for help in GLoBES.
\end{acknowledgments}

\end{document}